\UseRawInputEncoding
\documentclass[a4paper,11pt]{article}
\pdfoutput=1
\usepackage{jheppub}
\usepackage{caption,subcaption}
\usepackage{graphicx} % demo is just for the example
\usepackage{rotating,float,afterpage}
\usepackage{calrsfs}
\usepackage{slashed,cancel}
\usepackage{relsize}
\usepackage{amsmath,graphicx,epsfig,amssymb,multirow}
\RequirePackage{mathptmx,flushend}      % use Times fonts if available on your TeX system
\usepackage[12hr]{datetime} %to be removed finally

\newcommand{\FDF}{\left(\varphi^\dagger\overleftrightarrow{D}_\mu\varphi\right)}
\newcommand{\FDFI}{\left(\varphi^\dagger\overleftrightarrow{D}^I_\mu\varphi\right)}
\allowdisplaybreaks

\def\MGvATNLO{{\tt {\sc MadGraph5}\_aMC@NLO}}
\def\tb{\bar{t}}
\def\ttz{t\bar{t}Z}
\def\l{\left}
\def\r{\right}

\title{On two-body and three-body spin correlations  in leptonic $t\bar{t}Z$ production and  anomalous couplings  at the LHC}

\author{Rafiqul Rahaman}
\affiliation{Regional Centre for Accelerator-based Particle Physics, Harish-Chandra Research Institute,\\ A CI of Homi Bhabha National Institute,  Chhatnag Road, Jhunsi, Prayagraj 211019, India}
\emailAdd{rafiqulrahaman@hri.res.in}

\abstract{
	
We study the anomalous  $\ttz$ couplings in the $\ttz$ production in leptonic final state at the $13$ TeV LHC. We use the polarizations of top quarks and $Z$ boson, two-body and three-body spin correlations among the top quarks and $Z$ boson, and the cross section to probe the anomalous couplings. We estimate one parameter and simultaneous limits on the couplings of the effective vertex as well as the effective operators  for a set of luminosities $150$ fb$^{-1}$,  $300$ fb$^{-1}$, $1000$ fb$^{-1}$, and $3000$ fb$^{-1}$. The polarizations and the spin correlations are found to be helpful on top of the cross section to better constrain the anomalous couplings.
}

\preprint{HRI-RECAPP-2022-006}
\begin{document}
	\maketitle	
	%%%%%%%%%%%%%%%%%
	%%%%%%%%%%%%%%%%%%%%%%%%%%%%%%%%%%%%%%%%%%%%%%%%%%%%%%%%%%%%%%%%%%%%%%%%%%%%%%%%%%%%%%%%%%%%%%%%%
	\section{Introduction}\label{sec:intro}
	The standard model (SM) of particle physics, though well established, requires corrections not only due to incompatibility with the experimental evidence of non zero neutrino mass, dark matter, and Baryogenesis but also the theoretical issues such as the hierarchy of mass scales, the strong $CP$ problem, etc. Anomalies in recent experiments such as  the muon $(g-2)$ anomaly~\cite{Muong-2:2021ojo}, and $W$-mass anomalies~\cite{CDF:2022hxs}, along with few fluctuations~\cite{ATLAS:2016gzy,CMS:2016xbb,CMS:2018lce,CMS:2019efc}, have strengthened the requirement to go beyond the SM (BSM), although no direct evidence has been observed for any BSM degrees of freedom at experiments. Precision measurements at the electroweak (EW) scale are, thus, essential to look for the remnant of BSM physics sitting at a high energy scale. Top quark interactions play a key role in exploring BSM theories because of their heavy mass, which is the same order as the electroweak scale. Because of the almost bare nature of the top quark, i.e., it decays before hadronization, observables associated with or induced by its spins are useful tools in the precision measurement of the top quark interaction.
	
	We are interested in the precise measurement of the top quark interaction with the $Z$ boson in the $\ttz$ production process, as it is the most sensitive process for the direct measurement of such interaction. The $\ttz$ process is important also because of the key background to several searches of BSM phenomena in multi-lepton and $b$-quark final state. The $\ttz$ process is also an important background for the $t\tb$ production associated with a Higgs, a key process to study the top quark and Higgs interaction. The polarizations of top quarks, $Z$ boson, and spin correlations of $t$, $\tb$, and $Z$ can be useful observables to study the $\ttz$ interaction. There has been a lot of interest in recent days in the polarizations of top quarks~\cite{Kane:1991bg,Jezabek:1994zv,Hikasa:1999wy,Godbole:2006tq,Perelstein:2008zt,Huitu:2010ad,Choudhury:2010cd,Arai:2010ci,Gopalakrishna:2010xm,Godbole:2010kr,Godbole:2011vw,Krohn:2011tw,Rindani:2011pk,Cao:2011hr,Rindani:2011gt,Bhattacherjee:2012ir,Fajfer:2012si,Biswal:2012dr,Belanger:2013gha,Baumgart:2013yra,Godbole:2015bda,Rindani:2015vya,Rindani:2015dom,Behera:2018ryv,Arhrib:2018bxc,Wu:2018xiz,Zhou:2019alr,Arhrib:2019tkr,Patrick:2019nhv}, $Z/W$ boson~\cite{Abbiendi:2000ei,Rahaman:2016pqj,Rahaman:2017qql,Rahaman:2017aab,Nakamura:2017ihk,Aguilar-Saavedra:2017zkn,Rahaman:2018ujg,Rao:2018abz,Renard:2018tae,Renard:2018bsp,Renard:2018lqv,Rahaman:2019mnz,Rahaman:2019lab,Rahaman:2020jll}, including the $t\tb$ spin correlation~\cite{Cheung:1996kc,Arai:2004yd,Arai:2007ts,Arai:2009cp,Yue:2010hy,Degrande:2010kt,Cao:2010nw,Baumgart:2011wk,Barger:2011pu,Fajfer:2012si,Kiers:2014uqa,Bernreuther:2013aga,Bernreuther:2015yna,Aguilar-Saavedra:2018ggp,Ravina:2021kpr,Cakir:2022plk} for the study of new physics beyond the SM. Measurement on the top polarizations and $t\tb$ spin correlations are performed earlier at the Tevatron~\cite{Aaltonen:2010nz,Abazov:2011ka,Abazov:2011qu,Abazov:2011gi,Abazov:2012oxa,Abazov:2015psg,Lee:2018lgy} and recently at the Large Hadron Collider (LHC) by ATLAS and CMS experiments~\cite{ATLAS:2012ao,Aad:2013ksa,Aad:2014pwa,Aad:2014mfk,Khachatryan:2015tzo,Aad:2015bfa,Tiko:2016brs,Khachatryan:2016xws,Aaboud:2016bit,CMS:2018jcg,Aaboud:2019hwz}. The $Z$ boson polarizations are also measured at the LHC recently~\cite{Aaboud:2019gxl,CMS:2021lix}. A recent measurement by ATLAS~\cite{Aaboud:2019hwz} shows a deviation on the $t\tb$ spin correlations compared to the SM prediction at next-to-leading (NLO) accuracy in QCD~\cite{Bernreuther:2010ny,Behring:2019iiv,Czakon:2020qbd,Frederix:2021zsh}. The $\ttz$ production process has been searched at the LHC by ATLAS~\cite{ATLAS:2019fwo} and CMS~\cite{CMS:2019too,CMS:2022hjj} experiments providing a consistent rate with the SM expectation~\cite{Frixione:2015zaa,LHCHCSWG:2016ypw,Broggio:2019ewu,Kulesza:2020nfh,Ghezzi:2021rpc,Bevilacqua:2019cvp,Bevilacqua:2022nrm}. These searches also probe the anomalous $\ttz$ interaction with the help of differential rates in various kinematic variables. The $\ttz$ anomalous couplings have also been studied theoretically~\cite{Baur:2004uw,Baur:2005wi,Berger:2009hi,Rontsch:2014cca,Rontsch:2015una,Schulze:2016qas,BessidskaiaBylund:2016jvp,HajiRaissi:2020eob,Cao:2020npb,Ravina:2021kpr} deriving constraints on them. 
	A recent study~\cite{Ravina:2021kpr} also discusses the possibilities of observing $t\tb$ spin correlations in the $\ttz$ production process, including the effect of anomalous $\ttz$ interaction.

	In this study, we use the polarizations of top quarks and $Z$ boson, spin correlations of $t\tb$ pair, $tZ$ pair, $\tb Z$ pair, including $t\tb Z$ triplet, and the total cross section to probe anomalous $\ttz$ interaction in the $\ttz$ production process at the $13$ TeV LHC. We work in the fully leptonic final state, i.e., in $4l+2b+\cancel{E}_T$ final state, for better reconstruction of three-body ($t$-$\bar{t}$-$Z$) spin correlation. We then extrapolate the result by combining all semi-leptonic channels. We use the effective vertex factor parameterization for the anomalous $\ttz$ interaction~\cite{Aguilar-Saavedra:2008nuh,AguilarSaavedra:2012vh,Ravina:2021kpr} to investigate them and translate their limit into the higher dimensional effective operators~\cite{Aguilar-Saavedra:2008nuh,BessidskaiaBylund:2016jvp,Ravina:2021kpr} to compare with the existing studies~\cite{CMS:2015uvn,Hartland:2019bjb,ATLAS:2019fwo,CMS:2019too,CMS:2022hjj}. The polarizations and spin correlations provide complementary sensitivity to the anomalous $\ttz$ interaction to the existing approaches. The use of a large number of polarization and spin correlation observables allows us to extract more information from a given measurement with low statistical correlation and may eventually help to isolate the contribution of individual interaction parameters.
	
	The rest of the article is organized as follows. In section~\ref{sec:spincorr-formlaism}, we present the formalism for the three-body spin correlations among $t$, $\tb$, and $Z$ and the method to obtain them in experiments or in a Monte-Carlo (MC) simulation. We discuss the signal process in section~\ref{sec:signal-events}, along with the effect of neutrino reconstruction on the polarizations and spin correlations in the SM. In section~\ref{sec:eft-effect}, we investigate the effect of anomalous $\ttz$ interaction on the observables, followed by extracting simultaneous limits on them. Finally, we summarize in section~\ref{sec:conclusion}.

	\section{Spin correlation formalism}\label{sec:spincorr-formlaism}
	The spin density matrix or the polarization correlation density matrix for  $t\bar{t}Z$ production, i.e., two spin-$1/2$ particles and one spin-$1$ particle, can be
	written as~\cite{Rahaman:2021fcz},
	\begin{eqnarray}\label{eq:spin-density-half-one}
		P^{\ttz}\l(\lambda_t,\lambda_t^\prime,\lambda_{\tb},\lambda_{\tb}^\prime,\lambda_Z,\lambda_Z^\prime\r) &=& \frac{1}{\l(2\times\frac{1}{2}+1\r)^2}\frac{1}{\l(2\times 1+ 1\r)}
		\Big[ \mathbb{I}_{12\times 12}  
		+ \vec{p}^t\cdot\vec{\tau}\otimes \mathbb{I}_{6\times 6}
		+\mathbb{I}_{2\times 2}\otimes\vec{p}^{\tb}\cdot\vec{\tau}\otimes \mathbb{I}_{3\times 3} \nonumber\\
		&+& \dfrac{3}{2}\mathbb{I}_{4\times 4} \otimes \vec{p}^Z\cdot\vec{S}
		+\sqrt{\dfrac{3}{2}} \mathbb{I}_{4\times 4} \otimes  T_{ij}^Z\big(S_iS_j+S_jS_i\big)    
		+ pp_{ij}^{t\tb}\tau_i\otimes \tau_j \otimes \mathbb{I}_{3\times 3}  \nonumber\\
		&+& pp_{ij}^{tZ}\tau_i\otimes\mathbb{I}_{2\times 2}\otimes S_j + pT_{ijk}^{tZ}\tau_i\otimes\mathbb{I}_{2\times 2}\otimes (S_jS_k+S_kS_j) \nonumber\\
		&+& pp_{ij}^{\tb Z}\mathbb{I}_{2\times 2}\otimes\tau_i\otimes S_j + pT_{ijk}^{\tb Z}\mathbb{I}_{2\times 2}\otimes \tau_i\otimes(S_jS_k+S_kS_j) \nonumber\\
		&+& ppp_{ijk}^{\ttz}\tau_i\otimes\tau_j\otimes S_k +ppT_{ijkl}^{\ttz}\tau_i\otimes\tau_j\otimes (S_kS_l+S_lS_k)  \Big],\nonumber\\&&~\l( i,j,k,l\in[ x\equiv 1,y\equiv 2,z\equiv 3] \r).
	\end{eqnarray} 
	Here, $\mathbb{I}_{n\times n}$ is the unit matrix in $n$-dimension; $\vec{p}^{t/\tb/Z}$ are the vector polarizations of $t/\tb/Z$; $T_{ij}^{Z}$ are the  tensor polarizations of $Z$; $pp^{AB}$ are the vector-vector spin correlations of $A$-$B$ ($A/B=t/\tb/Z$) pair;  $pT_{ijk}^{tZ}$ ($pT_{ijk}^{\tb Z}$) are the  vector-tensor spin correlations of $t$-$Z$ ($\tb$-$Z$) pair; $ppp_{ijk}^{\ttz}$ and $ppT_{ijkl}^{\ttz}$ are the vector-vector-vector  and vector-vector-tensor spin correlations, respectively of $t$-$\tb$-$Z$ system. The $pT_{i(jk)}^{(t/\bar{t})Z}$ and $ppT_{ij(kl)}^{\ttz}$ are symmetric in the last two indices similar to the $T^Z$~\cite{Rahaman:2021fcz}.  The independent polarizations and spin correlations are as follows. There are
	\begin{itemize}
		\item $p^{t/\tb/Z}$: nine ($3\times 3=9$) vector polarizations of $t$, $\bar{t}$, and $Z$, 
		\item  $T^Z$: five tensor polarizations of $Z$,
		\item $pp^{t\tb}$, $pp^{tZ}$, $pp^{\tb Z}$: twenty seven ($3\times 3\times 3=27$) vector-vector spin correlations of $t$-$\tb$, $t$-$Z$ and $\tb$-$Z$ pairs,
		\item $pT^{tZ}$, $pT^{\tb Z}$: thirty ($2\times 3\times 5=30$) vector-tensor spin correlations of $t$-$Z$ and $\tb$-$Z$ pairs,
		\item $ppp^{\ttz}$: twenty seven ($3\times 3\times 3=27$) vector-vector-vector spin correlations of $t$-$\tb$-$Z$ system, and 
		\item  $ppT^{\ttz}$: forty five ($3\times 3\times 5 = 45$) vector-vector-tensor spin correlations of $t$-$\tb$-$Z$ system. 
	\end{itemize}
	Thus, there are a total of $143$ polarization and spin correlation parameters in the $\ttz$ process. Here, $pp$ and $pT$ are the two-body spin correlations, while $ppp$ and $ppT$ are the three-body spin correlations.
	The joint  angular distribution of the leptons  will contain the three-body angular functions such as
	\begin{eqnarray}\label{eq:norm-ttz-ang-dist}
		\dfrac{1}{\sigma}\dfrac{d^3\sigma}{d\Omega_{l_t}d\Omega_{l_{\tb}}d\Omega_{l_Z}}&=&\dfrac{1}{64\pi^3}\Biggr[1+
		\alpha_t\alpha_{\tb} \alpha_Z ~ppp_{ijk}^{\ttz}~ c_{i}^{l_t} c_{j}^{l_{\tb}} c_k^{l_Z}\nonumber\\
		&+& \alpha_t\alpha_{\tb}(1-3\delta_Z)ppT_{ijkl}^{\ttz} c_{i}^{l_t} c_{j}^{l_{\tb}} c_{k}^{l_Z} c_{l}^{l_Z}~(k\ne l)\nonumber\\
		&+&\frac{1}{2}\alpha_t\alpha_{\tb} (1-3\delta_Z)\underbrace{\left(ppT_{ijxx}^{\ttz}-pT_{ijyy}^{\ttz}\right)}_{ppT_{ij(x^2-y^2)}^{\ttz}} c_{i}^{l_t} c_{j}^{l_{\tb}}\left((c_{x}^{l_Z})^2-(c_{y}^{l_Z})^2\right) \nonumber\\
		&+&\frac{1}{2} \alpha_t\alpha_{\tb} (1-3\delta_Z) ppT_{ijzz}^{\ttz} c_{i}^{l_t} c_{j}^{l_{\tb}} \l(3(c_{z}^{l_Z})^2-1\r)\Biggr]
	\end{eqnarray}
	apart from the two-body and single particle angular functions as described in Ref.~\cite{Rahaman:2021fcz} with $c_x$, $c_y$, and $c_z$ as the angular functions of the leptons, i.e.,
	\begin{eqnarray}\label{eq:cor-defination}
		c_{x}^l = \sin\theta_l\cos\phi_l,~c_{y}^l = \sin\theta_l\sin\phi_l,~c_{z}^l = \cos\theta_l .
	\end{eqnarray}
	Here, $l_A$ denotes lepton decayed from the particle $A$. In this case, $\alpha_t=1=-\alpha_{\tb}$, $\alpha_Z \simeq -0.22$ and $\delta_Z=0$ in the SM~\cite{Boudjema:2009fz}. 
	The three-body spin correlations can be obtained from the following asymmetries,
	\begin{eqnarray}\label{eq:asym-App-half-one}
		{\cal A}\left[ppp_{ijk}^{\ttz}\right] 
		&=&\dfrac{\sigma\l(c_i^{l_t} c_j^{l_{\tb}} c_k^{l_Z}>0\r)-\sigma\l(c_i^{l_t} c_j^{l_{\tb}} c_k^{l_Z}<0\r)}
		{\sigma\l(c_i^{l_t} c_j^{l_{\tb}} c_k^{l_Z}>0\r)+\sigma\l(c_i^{l_t} c_j^{l_{\tb}} c_k^{l_Z}<0\r)},\nonumber\\
		&=&\frac{1}{8}\alpha_t\alpha_{\tb} \alpha_Z ~ppp_{ijk}^{\ttz},\nonumber\\
		%\end{eqnarray}
		%\begin{eqnarray}\label{eq:asym-ApT-half-one}
		{\cal A}\left[ppT_{ij(kl)}^{AB}\right] 
		&=&\dfrac{\sigma\l(c_i^{l_t} c_j^{l_{\tb}} c_k^{l_Z} c_l^{l_Z}>0\r)-\sigma\l(c_i^{l_t} c_j^{l_{\tb}c_l^{l_Z}} c_k^{l_Z}<0\r)}
		{\sigma\l(c_i^{l_t} c_j^{l_{\tb}} c_k^{l_Z} c_l^{l_Z}>0\r)+\sigma\l(c_i^{l_t} c_j^{l_{\tb}c_l^{l_Z}} c_k^{l_Z}<0\r)},~(k\ne l), \nonumber\\
		&=& \frac{1}{6\pi}\alpha_t\alpha_{\tb}(1-3\delta_B) ppT_{ij(kl)}^{\ttz},\nonumber\\ 
		{\cal A}\left[ppT_{ij(x^2-y^2)}^{\ttz}\right] 
		&=&\dfrac{\sigma\l(c_i^{l_t} c_j^{l_{\tb}} \l((c_{x}^{l_Z})^2-(c_{y}^{l_Z})^2\r) >0\r)-\sigma\l(c_i^{l_t} c_j^{l_{\tb}} \l((c_{x}^{l_Z})^2-(c_{y}^{l_Z})^2\r) <0\r)}{\sigma\l(c_i^{l_t} c_j^{l_{\tb}} \l((c_{x}^{l_Z})^2-(c_{y}^{l_Z})^2\r) >0\r)+\sigma\l(c_i^{l_t} c_j^{l_{\tb}} \l((c_{x}^{l_Z})^2-(c_{y}^{l_Z})^2\r) <0\r)},\nonumber\\
		&=& \frac{1}{6\pi}\alpha_t\alpha_{\tb}(1-3\delta_B) ppT_{ij(x^2-y^2)}^{\ttz},\nonumber\\ 
		{\cal A}\left[ppT_{ijzz}^{\ttz}\right] 
		&=& \dfrac{\sigma\l(c_i^{l_t} c_j^{l_{\tb}} \sin\l(3\theta_{l_Z}\r)>0\r)-\sigma\l(c_i^{l_t} c_j^{l_{\tb}} \sin\l(3\theta_{l_Z}\r)<0\r)}{\sigma\l(c_i^{l_t} c_j^{l_{\tb}} \sin\l(3\theta_{l_Z}\r)>0\r)+\sigma\l(c_i^{l_t} c_j^{l_{\tb}} \sin\l(3\theta_{l_Z}\r)<0\r)}  ,\nonumber\\
		&=& \frac{3}{32}\alpha_t\alpha_{\tb}(1-3\delta_B) ppT_{ijzz}^{\ttz}.
	\end{eqnarray}
	Asymmetries for the polarizations of $t$, $\tb$ and $Z$, and the spin correlations for $t$-$\tb$ and $t/\tb$-$Z$ are described in Ref.~\cite{Rahaman:2021fcz}. These formulas  can be used to estimate the 
	three-body spin correlations with real data from experiments or with events generated by a Monte-Carlo simulation. In the next section, we calculate these polarization and spin correlation parameters from their asymmetries with events generated at \MGvATNLO~\cite{Alwall:2014hca} after reconstructing the two missing neutrinos.

	\section{Signal process, polarizations and spin correlations in the SM}\label{sec:signal-events}
	\begin{figure}[h!]
		\begin{center}
			\includegraphics[width=1\textwidth]{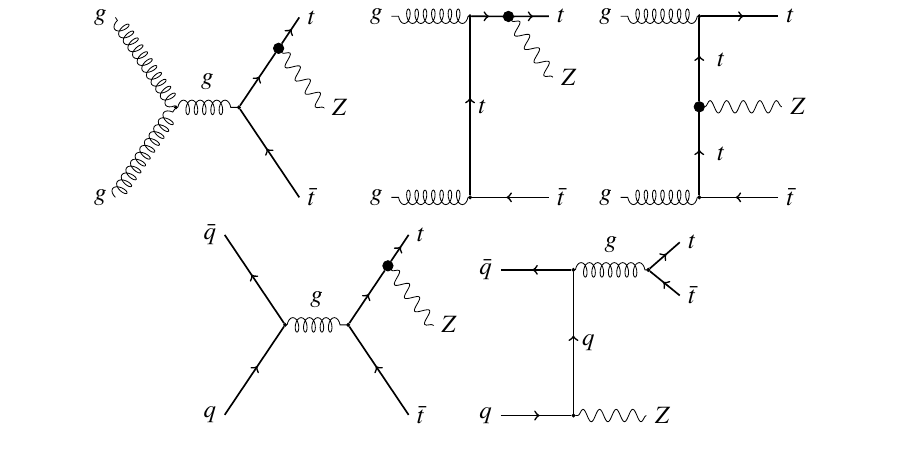}
		\end{center}
		\caption{\label{fig:Feynman-ttz} Representative Feynman diagrams for $\ttz$ production at leading order at the LHC. The shaded blob presents the presence of anomalous $\ttz$ couplings.} 
	\end{figure}
	We are interested in the signal topology of $\ttz$ production in the fully leptonic decay channel  at the $13$ TeV LHC. The $\ttz$ process comprises $gg$ and $q\bar{q}$ initial beams coming from proton as partons, see Fig.~\ref{fig:Feynman-ttz} for the representative Feynman diagrams at leading order (LO), where a $Z$ radiation from $\bar{t}$ leg is also assumed implicitly. The production cross section of the $\ttz$ process is estimated to be $\sigma_{\ttz}^{\text{LO}} \simeq 0.59$ fb at LO for $\sqrt{s}=13$ TeV with the package \MGvATNLO~v2.7.3, while the experimentally measured cross section by CMS~\cite{CMS:2019too} is $0.95\pm0.05_{\text{stat.}}+0.06_{\text{syst.}}$, much larger than the LO estimate. We thus estimate the $\ttz$ production cross section at next-to-leading order (NLO) in QCD in \MGvATNLO~v2.7.3; the NLO cross section is $\sigma_{\ttz}^{\text{NLO}}\simeq 0.86$. Thus an NLO to LO factor of $\kappa_{\text{NLO}}=1.46$ will be used for the SM process including the decay later on~\cite{Frixione:2015zaa}. We use {\tt NNPDF31}~\cite{NNPDF:2017mvq} set for the parton distribution functions (PDFs) with $\alpha_s(m_Z)=0.118$. A fixed renormalization ($\mu_R$) and 
	factorization ($\mu_F$) of $\mu_F=\mu_R=m_t+m_Z/2$ is used along with the following SM input parameters~\cite{ParticleDataGroup:2018ovx}:
	\begin{eqnarray}\label{eq:sm-inputs}
		&m_t=173.0~\text{GeV},~~~ &\Gamma_t=1.508 ~\text{GeV},   \\
		&m_Z=91.118~\text{GeV},~&\Gamma_W=2.085 ~\text{GeV}, \\
		&m_W=80.42~\text{GeV},~~~&\Gamma_Z=2.4952 ~\text{GeV},\\
		&m_b=4.7~\text{GeV}, ~~~~~~~&G_F=1.16637\times 10^{-5}~\text{GeV}^{-2}.
	\end{eqnarray}
	
	Signal events ($100$ million) for the $\ttz$ production and their leptonic decays are generated in \MGvATNLO~v2.7.3 at LO  with  generation level cuts  of 
	\begin{eqnarray}\label{eq:gen-cut}
		&p_T(b)>20~\text{GeV},~p_T(l)>10~\text{GeV},~|\eta_b|<5.0,~|\eta_l|<2.5,\nonumber\\
		&\Delta R(b,b)>0.4,~~\Delta R(b,l)>0.4,~\Delta R(l,l)>0.02
	\end{eqnarray}
	with the input parameters discussed above. 
	We generated the events in \MGvATNLO~as
	\begin{equation}\label{eq:process}
		pp\to \ttz,~ t\to b (W^+\to \mu^+ \nu_\mu),~ \bar{t}\to \bar{b} (W^-\to \mu^- \bar{\nu}_\mu),~ Z\to e^-e^+,
	\end{equation}
	i.e., tops are decayed in muonic flavors, while the $Z$ boson is decayed in electronic flavor for simplicity. A flavor factor of $8$ for the complete leptonic channel is accounted for in the analysis.
	The parton level  events are then passed to {\tt PYTHIA8}~\cite{Sjostrand:2014zea} for showering and hadronization followed by fast detector simulation by {\tt Delphes}-3.5.0~\cite{deFavereau:2013fsa}. 
	The events are selected at the {\tt Delphes} level with at least two oppositely charged muons, two oppositely charged electrons, and two $b$-tagged jets using  
	the default isolation criteria,  $p_T$ and $\eta$ cuts given by
	\begin{eqnarray}\label{eq:sel-cut}
		&p_T(b)>20~\text{GeV},~p_T(l)>10~\text{GeV},~|\eta_b|<2.5,~|\eta_l|<2.5,\nonumber\\
		&%\cancel{E}_T>10~\text{GeV},~
		R_0(j)=0.5,~\Delta R_{\text{max}}(l)=0.5.
	\end{eqnarray}
	The detection efficiency of the $2b+2\mu+2e+\cancel{E}_T$ events with the above selection criteria is about $\epsilon_{sel.}\simeq10.2\%$. The estimated number of events in a fully leptonic channel, i.e., including the flavor factor, adjusted with an NLO $k$-factor of $1.46$, after the selection criteria is about $N_{\text{SM}}=403$ for an integrated luminosity of ${\cal L}=3000$ fb$^{-1}$. 
	These events are then used to calculate all the polarizations and spin correlations and their asymmetries after reconstructing the two missing neutrinos needed to obtain the top quarks' rest frame. The method for reconstruction of the two neutrinos is described in the following section, followed by comparing the reconstructed variables to the truth level variables. 
	
	\subsection{Reconstruction of the neutrinos }\label{sec:neutrino-reco}	
	\begin{figure}[h!]
		\begin{center}
			\includegraphics[width=0.49\textwidth]{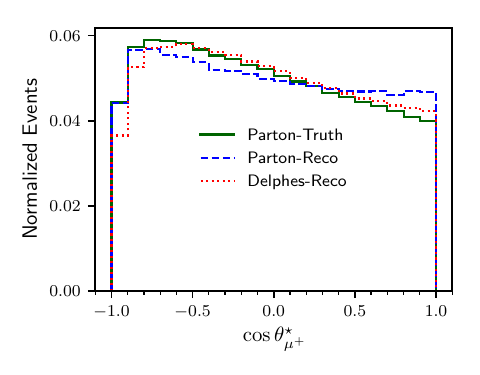}
			\includegraphics[width=0.49\textwidth]{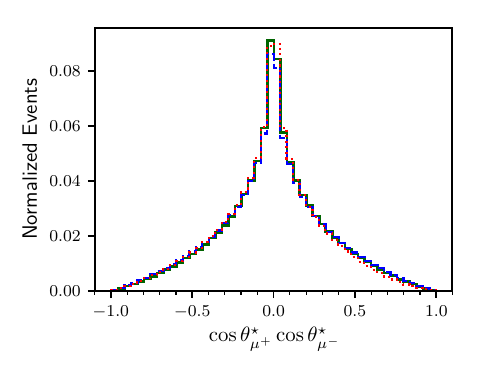}
			\includegraphics[width=0.49\textwidth]{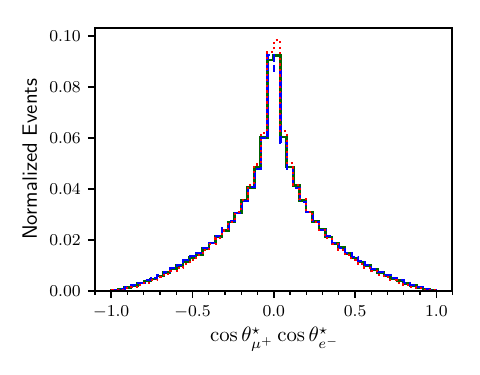}
			\includegraphics[width=0.49\textwidth]{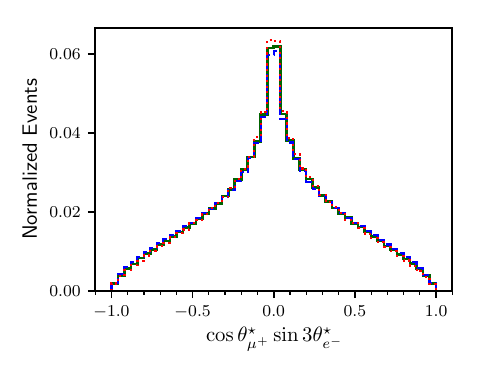}
		\end{center}
		\caption{\label{fig:pol-cor-2body} Normalized distributions of angular functions for polarizations and two-body spin correlations ($p_z^t$ in {\em left-top}, $pp_{zz}^{t\bar{t}}$ in {\em right-top}, $pp_{zz}^{tZ}$ in {\em left-bottom}, and $pT_{z(zz)}^{tZ}$ in {\em right-bottom} panel) in the rest frame of $t$, $\bar{t}$ and $Z$ for three scenarios namely {\tt Parton-Truth}, {\tt Parton-Reco} and {\tt Delphes-Reco}.} 
	\end{figure}
	\begin{figure}
		\begin{center}
			\includegraphics[width=0.49\textwidth]{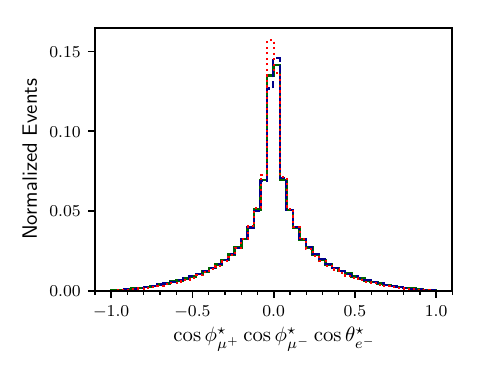}
			\includegraphics[width=0.49\textwidth]{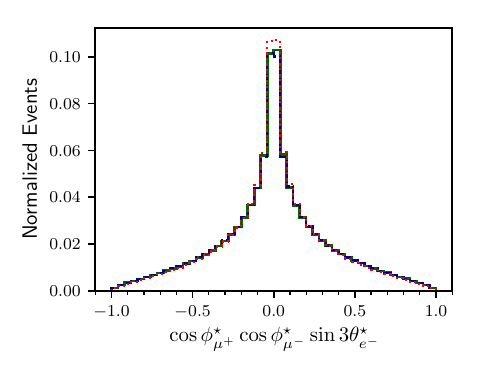}
			\includegraphics[width=0.49\textwidth]{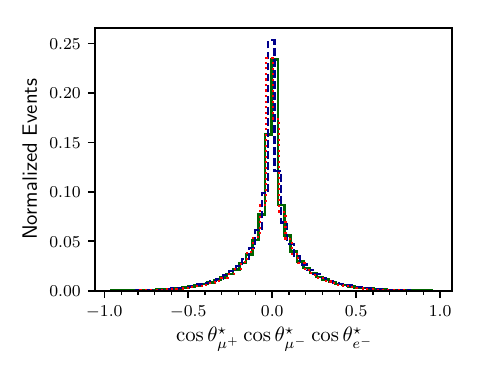}
			\includegraphics[width=0.49\textwidth]{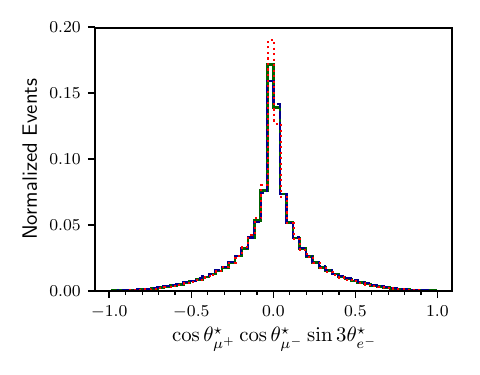}
		\end{center}
		\caption{\label{fig:cor-3body} Normalized distributions of angular functions for three-body ($t\bar{t}Z$)  spin correlations ($ppT_{xxz}^{t\bar{t}Z}$ in {\em left-top}, $ppT_{xx(zz)}^{t\bar{t}Z}$ in {\em right-top}, $ppT_{zzz}^{t\bar{t}Z}$ in {\em left-bottom}, and $ppT_{zz(zz)}^{t\bar{t}Z}$ in {\em right-bottom} panel) in the rest frame of $t$, $\bar{t}$ and $Z$ for three scenarios namely {\tt Parton-Truth}, {\tt Parton-Reco} and {\tt Delphes-Reco} with the legends same as in Fig.~\ref{fig:pol-cor-2body}.} 
		
	\end{figure}
	We reconstruct the two neutrinos with detector level events for the polarizations and spin correlations in the $\ttz$ process in the decay channel given in Eq.~(\ref{eq:process}) as follows~\cite{CMS:2015rld,CMS:2018adi,CMS:2018jcg}.  
	At first, the two $b$-quarks are assigned with the correct lepton (here muon) such that
	$m_{b\mu^+}^2 + m_{\bar{b}\mu^-}^2 $ is minimum.
	The  four momenta of the two neutrinos are then solved using the  constraints,
	\begin{eqnarray}\label{eq:neu-sol-constraints}
		&\vec{\cancel{p}}_T =  \vec{p}_T(\nu_\mu) + \vec{p}_T(\bar{\nu}_\mu),&\nonumber\\
		&m_{\mu^+ \nu_\mu}^2=m_W^2=m_{\mu^- \bar{\nu}_\mu}^2,&\nonumber\\
		&m_{b\mu^+ \nu_\mu}^2=m_t^2=m_{\bar{b}\mu^- \bar{\nu}_\mu}^2.
		% &min\{m_{t\bar{t}}\}.&
	\end{eqnarray}
	These constraints provide multiple sets of solutions for the components of neutrinos' momenta consisting of real as well as imaginary values due to the quadratic nature of the constraint equations.
	In case all solutions are imaginary, the values of $m_W$ and $m_t$ are varied  $1000$ times following the Gaussian distribution with mean at $m_W$ and $m_t$ and variance  $10$ GeV and $40$ GeV, respectively, until at least one set of real solution for neutrinos momenta is obtained. The events are rejected if no set of real solutions is found within the iteration of $1000$, which is only about $0.2\%$. In an event consisting of several real sets of solutions for neutrinos momenta, the set is chosen with minimum $m_{t\bar{t}}$. 
	
	We study the goodness of the reconstruction method of the neutrinos in the SM by comparing the detector level results with the truth level results for the angular variables related to polarizations and spin correlations of $t$, $\bar{t}$ and $Z$, as described in section~\ref{sec:spincorr-formlaism}. Normalized distributions for some variables are shown in Fig.~\ref{fig:pol-cor-2body} for the polarizations and two-body spin correlations of the top quark and $Z$ boson, and in Fig.~\ref{fig:cor-3body} for three-body spin correlations as representative for the following three cases. 
	\paragraph*{I. {\tt Parton-Truth}:} In this method, observables are calculated with the parton level events with MC truth information of the final state particles, including the neutrinos.
	\paragraph*{II. {\tt Parton-Reco}:} In this method, we use the parton level events, but the two neutrinos are reconstructed using truth level pairing of $b$-quarks. 
	\paragraph*{III. {\tt Delphes-Reco}:} Here, we use the events selected at the {\tt Delphes} detector simulations to reconstruct the neutrinos using the above mentioned approach.
	
	\paragraph*{}
	The distributions for the polarizations of the top quark seem to be distorted in {\tt Parton-Reco} compared to the {\tt Parton-Truth} case, see Fig.~\ref{fig:pol-cor-2body} {\em left-top} panel. However, the final distributions in {\tt Delphes-Reco} become very similar to the {\tt Parton-Truth} because of events migration in detector simulation. Not much of a difference can be seen for the two-body spin correlations (see Fig.~\ref{fig:pol-cor-2body} excluding {\em left-top} panel), while a little difference can be seen for the three-body spin correlations in  {\tt Delphes-Reco} compared to {\tt Parton-Truth} (see Fig.~\ref{fig:cor-3body}). We also calculated the values for all the polarizations and spin correlations and their respective asymmetries in the SM. The values of various polarizations and spin correlations along with their asymmetries in parton level as well as in {\tt Delphes} level  are listed in Table~\ref{app:SM-pol-asym} in appendix~\ref{app:SM-pol-asym} for completeness. Though we can recover some of the polarization and spin correlation variables with good accuracy, we do not obtain a good accuracy for many variables after reconstructing the neutrinos at the detector level simulations.
	Nevertheless, for the sake of realistic analysis, we will calculate the polarization and spin correlation asymmetries at the {\tt Delphes} level by reconstructing the missing neutrinos with the above mentioned method  with anomalous couplings and higher dimensional operators. We investigate the new physics effect on the polarizations and spin correlations in the following section.
	
	\paragraph{$Z$ boson reconstruction and reference $z$-axis}
	One needs to identify the leptons coming from $Z$ in order to reconstruct the $Z$ boson to evaluate its polarizations and spin correlations with the top quarks. The $3e+\mu$ and $3\mu+e$ channels suffer a two-fold ambiguity, while the  $4\mu$ and $4e$ channels suffer four fold-ambiguity for the identification of $Z$ candidate leptons. Nevertheless, by demanding a same flavor oppositely charged lepton pair with an invariant mass closest to $m_Z$, these ambiguities can be resolved with reasonable accuracy. We use the $2e+2\mu$ channel as the proxy for the $4l$ final state for simplicity. We need a reference $z$-axis and $x$-$z$ plane to measure the angular orientations of the leptons for the polarizations and spin correlations. We consider the direction of the reconstructed boost as the proxy for the positive $z$-axis. The $t$-$\tb$ production plane can be safely considered as the $x$-$z$ plane or the $\phi=0$ plane to estimate polarizations and spin correlations.

	\section{Probe of anomalous $\ttz$ interaction}\label{sec:eft-effect}
	In this work, we are interested in studying the anomalous interactions, i.e., a contribution that can be received from beyond the SM, of  $t$, $\bar{t}$, and $Z$ in the $\ttz$ production process. We neglect the effect of the four-point contact interaction of four quarks that can enter into the $\ttz$ production at the LHC, as these are better constrained in $t\bar{t}$ production process. 
	The $\ttz$ interaction Lagrangian, including new physics, is generally parameterized in a model independent way as~\cite{Aguilar-Saavedra:2008nuh,AguilarSaavedra:2012vh,Ravina:2021kpr},
	\begin{equation}\label{eq:BSM-lag}
		{\cal L}_{\ttz} = e\bar{t}\left[ \gamma^\mu\left(C_1^V+\gamma_5 C_1^A\right) + \frac{i \sigma^{\mu\nu} q_\nu}{m_Z}\left( C_2^V+i\gamma_5 C_2^A \right) \right]t Z_\mu,
	\end{equation}
	with $\sigma^{\mu\nu}=\frac{1}{2}\left[\gamma^\mu,\gamma^\nu \right]$ and $q_\nu=\left(p_t-p_{\bar{t}}\right)_\nu$, i.e., four momentum transfer of $Z$, with $p_t$ ($p_{\bar{t}}$) as the four momentum of top (anti-top) quark. In the SM, $C_2^{V/A}=0$ at tree level and $C_1^{V/A}$ attain their SM values, which are $C_{1,\text{SM}}^V\simeq 0.24$ and $C_{1,\text{SM}}^A\simeq-0.60$. The couplings $C_2^V$ and $C_2^A$ ($CP$-odd) are the source of weak magnetic and electric
	dipole moments of the top quark, respectively and are highly suppressed in the SM~\cite{Rontsch:2014cca,Schulze:2016qas,Aguilar-Saavedra:2008nuh}.

	In the effective field theory (EFT) approach with higher dimension operators (${\cal O}$), the interaction Lagrangian is expressed as,
	\begin{equation}
		{\cal L}_{\ttz} = {\cal L}_{\ttz}(\text{SM}) + \sum_n\sum_i\dfrac{C_i^n}{\Lambda^{(n-4)}} {\cal O}_i^{n}
	\end{equation}
	with index $n$  summing over higher dimensions ($>4$) and index $i$ summing over all the operators in a givendimension; $C_i$ are the Wilson coefficients (WCs) corresponding to operator ${\cal O}_i$. The $\Lambda$ is the cut-off energy scale up to which the EFT is valid. The gauge invariant operators, made of the SM fields,  are low energy  remnant of some new physics theory at a higher energy scale ($>\Lambda$) with heavy fields, i.e., the heavy degrees of freedom are integrated out to the WCs at low energy. 
	The anomalous $\ttz$ interactions in Eq.~(\ref{eq:BSM-lag}) receive contribution from the following dimension-$6$ operators~\cite{Aguilar-Saavedra:2008nuh,BessidskaiaBylund:2016jvp,Ravina:2021kpr},
	\begin{eqnarray}\label{eq:BSM-Op}
		{\cal O}_{uB}^{ij} &=& \left( \bar{Q}_i\sigma^{\mu\nu} u_j \right) \tilde{\varphi} B_{\mu\nu},\nonumber\\
		{\cal O}_{uW}^{ij} &=& \left( \bar{Q}_i\sigma^{\mu\nu}\tau^I u_j \right) \tilde{\varphi} W_{\mu\nu}^I,\nonumber\\
		{\cal O}_{\varphi u}^{ij} &=& \FDF \left( \bar{u_i} \gamma^\mu u_j \right),\nonumber\\
		{\cal O}_{\varphi Q}^{1(ij)} &=& \FDF \left( \bar{Q_i} \gamma^\mu Q_j \right),\nonumber\\
		{\cal O}_{\varphi Q}^{3(ij)} &=& \FDFI \left( \bar{Q_i} \gamma^\mu \tau^I Q_j \right).
	\end{eqnarray}
	Here, $i,j$ are the flavor indices;  $Q$ is the left-handed quark doublet and $u$ are the right-handed singlet quark; $\tau^I$ are the Pauli matrices, $\varphi$ is the Higgs doublet with its dual $\widetilde{\varphi}=i\tau^2\varphi^\star$; $B_{\mu\nu}^I$ and $W_{\mu\nu}^I$ are the $U(1)_Y$ and $SU(2)_L$ filed strength tensor, respectively; $D_\mu$ ($\overleftarrow{D}_\mu$) is the
	covariant derivative acting on the right (left) and $\overleftrightarrow{D}_\mu=D_\mu-\overleftarrow{D}_\mu$.
	The couplings in Eq.~(\ref{eq:BSM-lag}) are related to the WCs of the operators listed in Eq.~(\ref{eq:BSM-Op}) as~\cite{Aguilar-Saavedra:2008nuh,Aguilar-Saavedra:2018ksv,Ravina:2021kpr},
	\begin{eqnarray}\label{eq:op-to-lag}
		C_1^V &=& C_{1,\text{SM}}^V + \frac{v^2}{2\Lambda^2\sin\theta_W\cos\theta_W} \text{Re}\left[ -c_{\varphi t} -c_{\varphi Q}^- \right],\nonumber\\
		C_1^A &=& C_{1,\text{SM}}^A + \frac{v^2}{2\Lambda^2\sin\theta_W\cos\theta_W} \text{Re}\left[ -c_{\varphi t} +c_{\varphi Q}^- \right],\nonumber\\
		C_2^V &=& \frac{\sqrt{2}v^2}{2\Lambda^2\sin\theta_W\cos\theta_W} c_{tZ},\nonumber\\
		C_2^A &=& \frac{\sqrt{2}v^2}{2\Lambda^2\sin\theta_W\cos\theta_W} c_{tZ}^I
	\end{eqnarray}
	with 
	\begin{eqnarray}\label{eq:cop}
		c_{tZ} &=& \text{Re}[-\sin\theta_W C_{uB}^{33} + \cos\theta_W C_{uW}^{33}],\nonumber\\
		c_{tZ}^I &=& \text{Im}[-\sin\theta_W C_{uB}^{33} + \cos\theta_W C_{uW}^{33}],\nonumber\\
		c_{\varphi t} &=& C_{\varphi u}^{33},\nonumber \\
		c_{\varphi Q}^- &=& C_{\varphi Q}^{1(33)}-C_{\varphi Q}^{3(33)}.
	\end{eqnarray}
	Here, $\theta_W$ is the Weinberg mixing angle, and $v\simeq 246$ GeV is the vacuum expectation value (VEV) of the Higgs. 
	We define $\Delta C_1^{V/A} = C_1^{V/A} - C_{1,\text{SM}}^{V/A} $ and treat  them as anomalous couplings in the rest of the article.
	
	\begin{figure}[ht!]
		\begin{center}
			\includegraphics[width=0.49\textwidth]{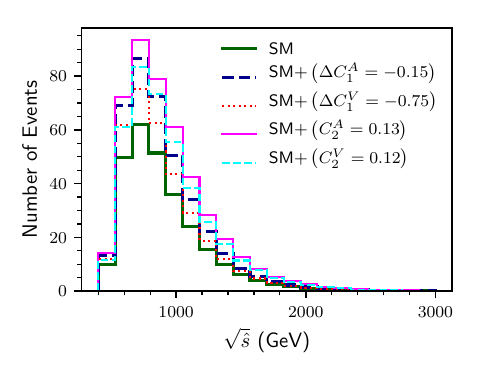}
			\includegraphics[width=0.49\textwidth]{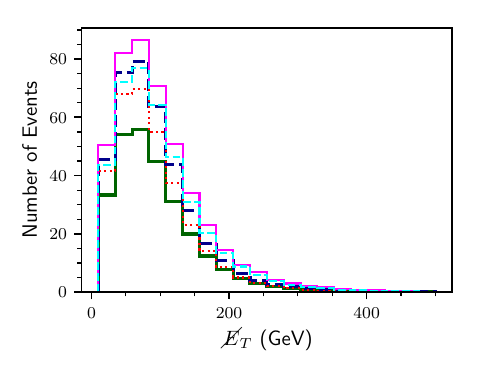}
			\includegraphics[width=0.49\textwidth]{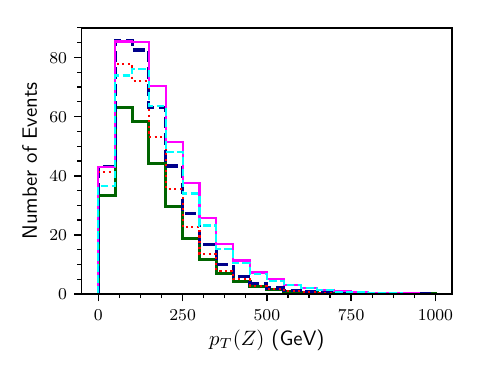}		
		\end{center}
		\caption{\label{fig:dist-kin} Distribution of $\sqrt{\hat{s}}$,  $\cancel{E}_T$, and $p_T(Z)$ in $2e+2\mu+2b+\cancel{E}_T$ final state for SM and few benchmark anomalous $t\bar{t}Z$ couplings with events normalized to an integrated luminosity of  ${\cal L}=3000$ fb$^{-1}$. }
	\end{figure}
	\begin{figure}[h!]
		\begin{center}
			\includegraphics[width=0.46\textwidth]{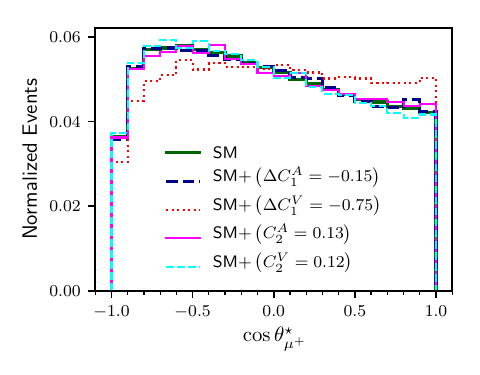}
				\includegraphics[width=0.46\textwidth]{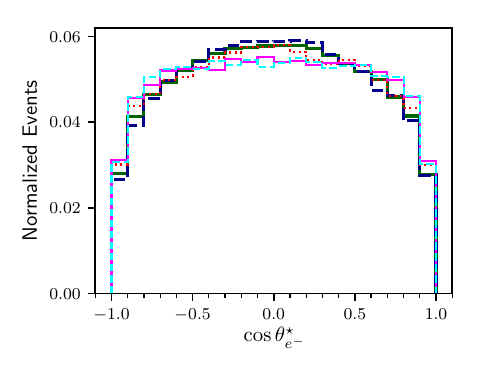}
			\includegraphics[width=0.46\textwidth]{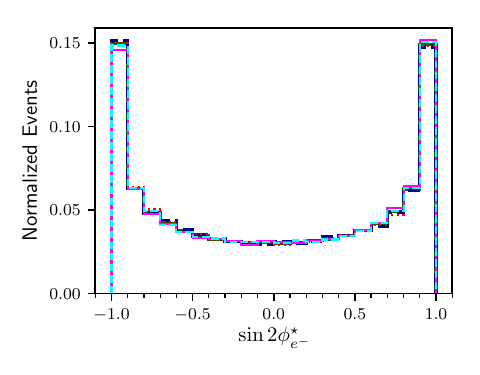}
			\includegraphics[width=0.46\textwidth]{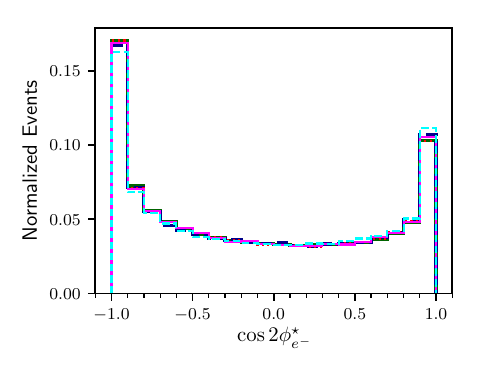}
		\end{center}
		\caption{\label{fig:dist-bsm-pol} Normalized distributions of angular functions for polarizations ($p_z^t$ in {\em left-top}, $p_z^Z$ in {\em right-top}, $T_{xy}^Z$ in {\em left-bottom}, and $T_{x^2-y^2}^Z$ in {\em right-bottom} panel) in the rest frame of $t$, $\bar{t}$ and $Z$ for SM and few benchmark anomalous $t\bar{t}Z$ couplings. } 
	\end{figure}
	\begin{figure}[h!]
		\begin{center}
			\includegraphics[width=0.46\textwidth]{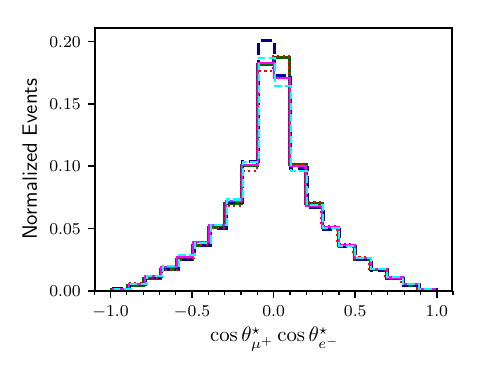}	
			\includegraphics[width=0.46\textwidth]{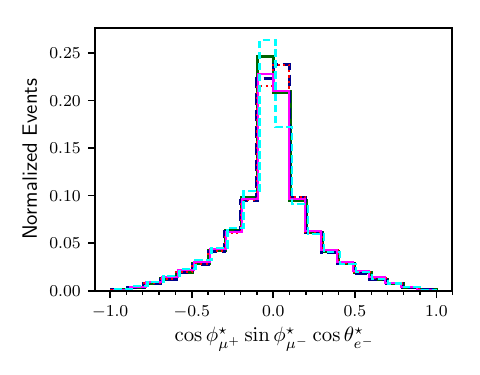}
			\includegraphics[width=0.46\textwidth]{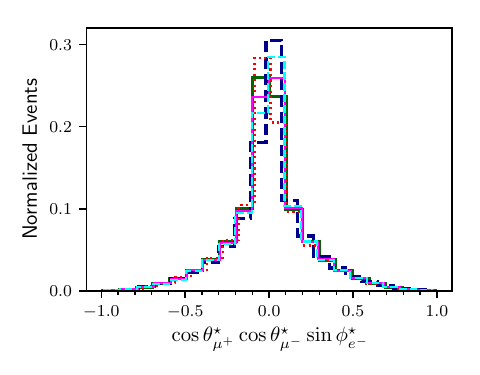}	
			\includegraphics[width=0.46\textwidth]{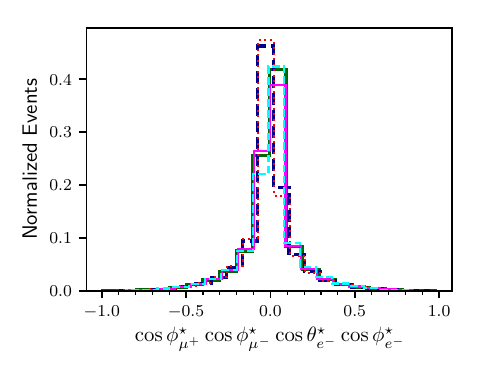}
		\end{center}
		\caption{\label{fig:dist-bsm-cor} Normalized distributions of angular functions for two-body and three-body spin correlations ( $pp_{zz}^{tZ}$ in {\em left-top}, $ppp_{xyz}^{\ttz}$ in {\em right-top}, $ppp_{zzy}^{\ttz}$ in {\em left-bottom}, and   $ppT_{xx(xz)}^{\ttz}$ in {\em right-bottom} panel) in the rest frame of $t$, $\bar{t}$ and $Z$ for SM and few benchmark anomalous $t\bar{t}Z$ couplings with the legends same as in Fig.~\ref{fig:dist-bsm-pol}. } 
	\end{figure}
	With the setup of anomalous $\ttz$ couplings and EFT operators discussed above, we proceed to probe them in our desired channel of $4l+2b+\cancel{E}_T$ final state with the polarization and spin correlation asymmetries along with the total rate (expected number of events with selection level cuts). We generated events ($10$ million) in \MGvATNLO~ of the process in Eq.~(\ref{eq:process}) as a proxy for the $4l+2b+\cancel{E}_T$ final state  for a set of  anomalous couplings  in Eq.~(\ref{eq:BSM-lag}) validated with {\tt FeynRules}~\cite{Alloul:2013bka} for the UFO model with the same set up of input parameters (SM inputs in Eq.~(\ref{eq:sm-inputs}), $\mu_F$, $\mu_R$, and PDFs)  and generation level cuts (Eq.~(\ref{eq:gen-cut})) as discussed in Section~\ref{sec:signal-events}. The parton level events are then passed through {\tt PYTHIA8} for showering and hadronization followed by detector simulation in {\tt Delphes}.  We, first, study the effect of the anomalous  couplings on the distributions of 
		kinematic variables such as the center-of-mass energy ($\sqrt{\hat{s}}$), after the reconstruction of neutrino momentum, discussed in Section~\ref{sec:neutrino-reco},  missing transverse energy ($\cancel{E}_T$), and transverse momentum of reconstructed $Z$ boson ($p_T(Z)$). The distributions are shown in Fig.~\ref{fig:dist-kin} with events normalized to an integrated luminosity of  ${\cal L}=3000$ fb$^{-1}$ for four benchmark anomalous couplings of $\Delta C_1^A=-0.15$, $\Delta C_1^V=-0.75$, $C_2^A=0.13$, and $C_2^V=0.12$ on top of the SM. 
	The $\sqrt{\hat{s}}$ distributions peak at around $700$ GeV for all the benchmark points with excess events in each bin compared to the SM case. The couplings $C_2^{V/A}$, being associated with momentum transfer of $Z$ (see Eq.~(\ref{eq:BSM-lag})), proide higher excess in events after the peak compared to $\Delta C_1^{V/A}$  with no momentum dependence. The excess in events is larger at the peak, and 
	thus no rectangular cuts on the $\sqrt{\hat{s}}$ will be able to enhance the signal to background ratios. The case is true for $\cancel{E}_T$ and $p_T(Z)$ along with the $p_T$s of all the particles, although not shown. We, thus, do not impose any rectangular cuts on the kinematic variables other than the selection cuts for our analysis; this will also help us not to diminish the statistics for the asymmetries. However, we use four bins in the $p_T(Z)$ for the cross section using the fact that $C_2^{V/A}$ shows excess in events for higher momentum transfer. The four bins in $p_T(Z)$ are chosen as 
		\begin{eqnarray}\label{eq:sigma-bin}
			Bin_1 &\equiv& p_T(Z) < 250 ~\text{GeV},\nonumber\\
			Bin_2 &\equiv& p_T(Z) \in[ 250,500 ]~\text{GeV},\nonumber\\
			Bin_3 &\equiv& p_T(Z) \in[ 500,750 ]~\text{GeV},\nonumber\\
			Bin_4 &\equiv& p_T(Z) > 750~\text{GeV}.
		\end{eqnarray}
		We will be using the same NLO to LO $k$-factor  of $1.46$ for the four bins in $p_T(Z)$, although it is higher for higher $p_T(Z)$~\cite{Frixione:2015zaa} . 
        For the asymmetries, however, we  use the un-binned cross section to avoid losing the statistics.

	%\paragraph{Distribution of polarization and spin correlation variables}
	We now investigate the effect of anomalous couplings on the angular distributions corresponding to polarizations and spin correlations, which are  shown in Fig.~\ref{fig:dist-bsm-pol} and Fig.~\ref{fig:dist-bsm-cor}, respectively as representative with the same benchmarks for anomalous couplings as used for kinematic variables. The top quark polarization ($p_z^t$) (Fig.~\ref{fig:dist-bsm-pol} {\em left-top} panel) shows deviation only for the $\Delta C_1^V$ benchmark, while  $p_z^Z$ (Fig.~\ref{fig:dist-bsm-pol} {\em right-top} panel) shows deviation only for $C_2^{A/V}$. Thus evidence for $\Delta C_1^V$ and $C_2^{A/V}$ can be clearly identified by looking at $p_z^t$ and $p_z^Z$ variables, respectively. The tensor polarizations for $Z$ boson such as $T_{xy}^Z$ in {\em left-bottom} and $T_{x^2-y^2}^Z$ in {\em right-bottom} show deviations for $C_2^A$ and $C_2^V$ couplings, respectively. The effect of the anomalous coupling $\Delta C_1^A$ is seen prominent in the $pp_{zz}^{tZ}$ spin correlation shown in the {\em left-top} panel of Fig.~\ref{fig:dist-bsm-cor}. The three body spin correlations  shown in Fig.~\ref{fig:dist-bsm-cor} (excluding {\em left-top} panel) show visible deviation from the SM for all the anomalous benchmark points as representative of many such spin correlation variables which are not shown. 
	We use all the polarization and spin correlation variables to study the sensitivity to the anomalous couplings and estimate their limits in the following subsection.
	
	\subsection{Constraints on the anomalous couplings and the operators}
	\begin{figure}[ht!]
		\begin{center}
			\includegraphics[width=0.8\textwidth]{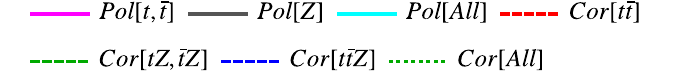}
				\includegraphics[width=0.49\textwidth]{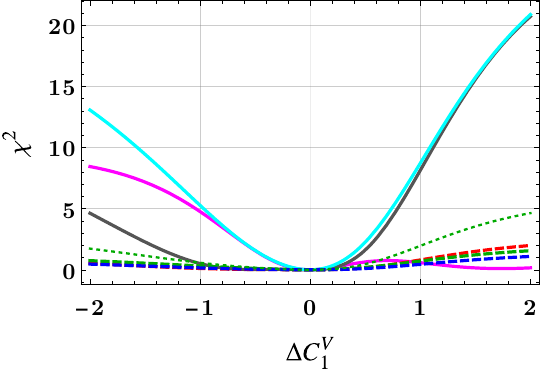}
				\includegraphics[width=0.49\textwidth]{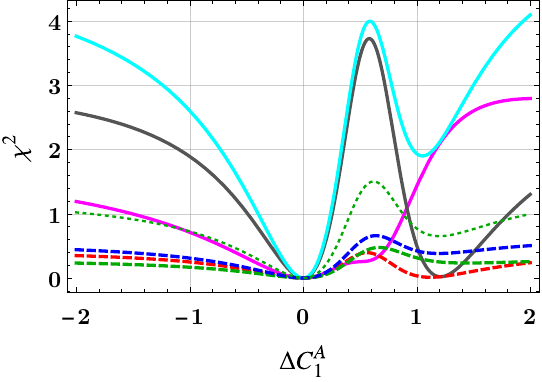}
			\includegraphics[width=0.49\textwidth]{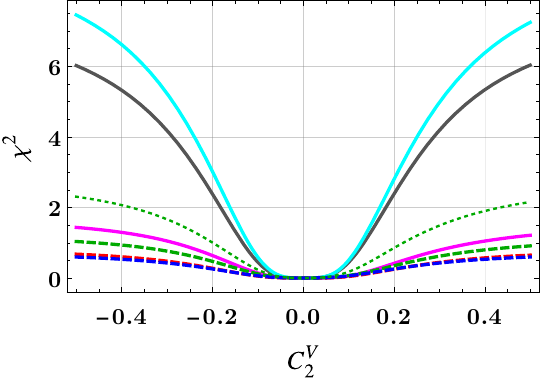}
			\includegraphics[width=0.49\textwidth]{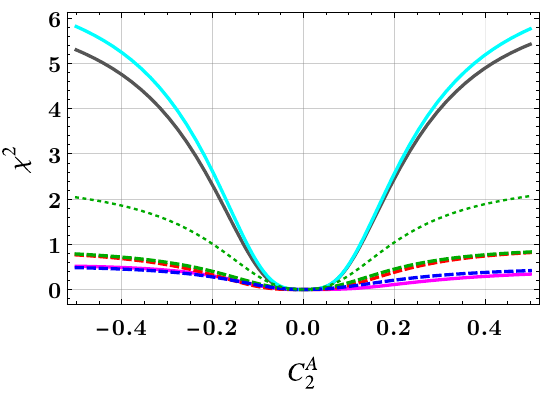}
		\end{center}
		\caption{\label{fig:bsm-asym} Comparison of different polarizations and spin correlations and their combinations in terms of $\chi^2$ as a function of $\ttz$ anomalous couplings at $\sqrt{s}=13$ TeV and integrated luminosity of  ${\cal L}=3000$ fb$^{-1}$.  } 
	\end{figure}
	%%%%%%%%%%%%%5
	\begin{figure}[ht!]
		\begin{center}
			\includegraphics[width=0.7\textwidth]{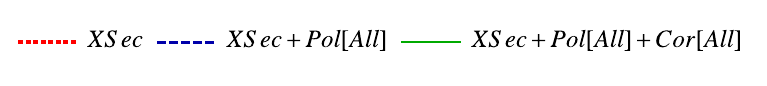}
			\includegraphics[width=0.45\textwidth]{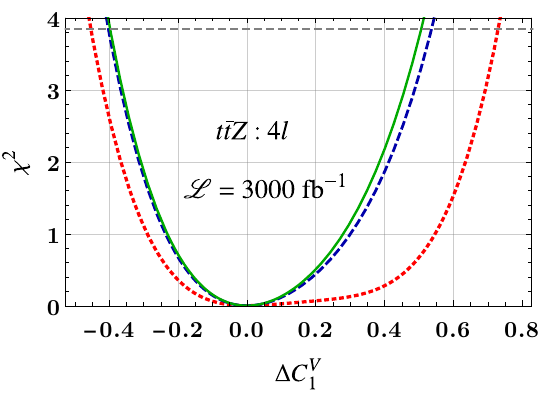}
			\includegraphics[width=0.45\textwidth]{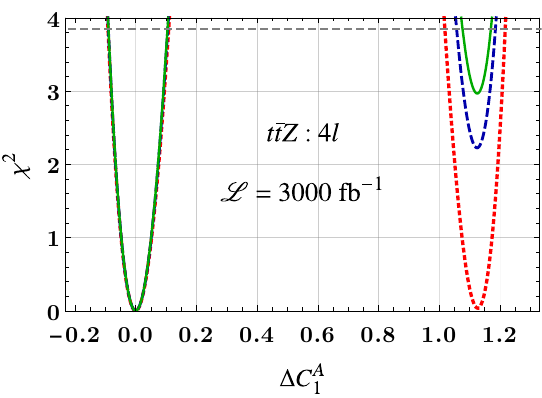}
			\includegraphics[width=0.45\textwidth]{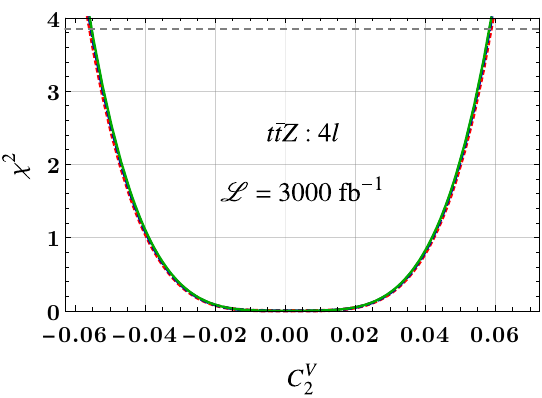}
			\includegraphics[width=0.45\textwidth]{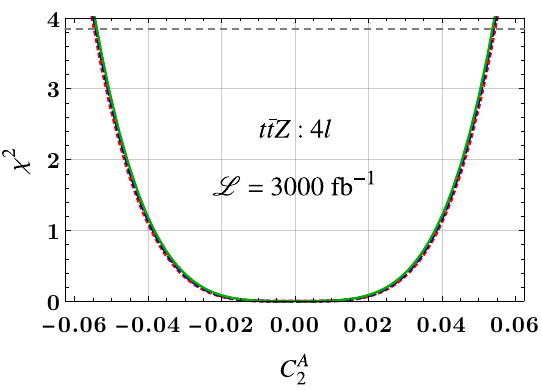}
			\includegraphics[width=0.42\textwidth]{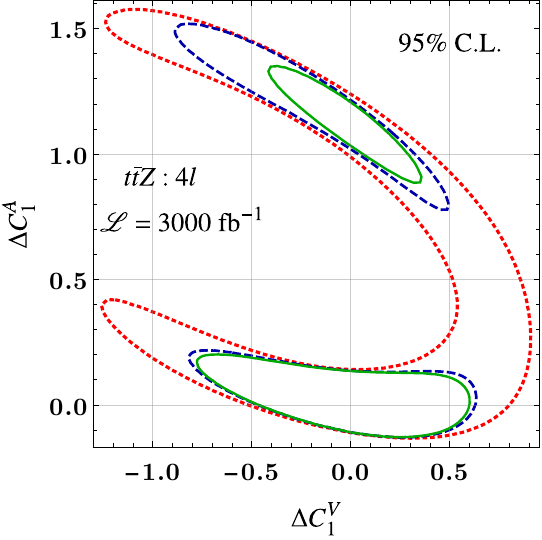}
			\includegraphics[width=0.45\textwidth]{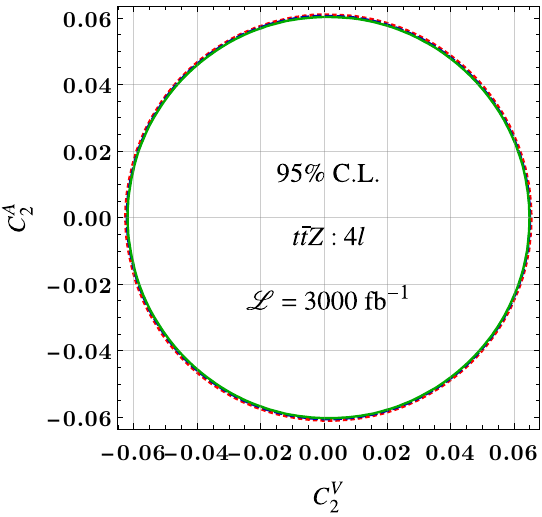}		
		\end{center}
		\caption{\label{fig:bsm-cs-polcor} The  $\chi^2$ for cross section ($XSec$), all polarizations ($Pol[All]$) and spin correlations  and their combinations are shown as a function of $\ttz$ anomalous couplings one at a time. The dashed horizontal lines at $\chi^2=3.84$ in the upper two rows indicate the $95\%$ C.L. bound on the couplings. The {\em bottom-panels} show the $95\%$ C.L contours for the above same combination of the observables. The plots are shown for $\sqrt{s}=13$ TeV and integrated luminosity ${\cal L}=3000$ fb$^{-1}$.  } 
	\end{figure}
	We use the {\tt Delphes} level events after selection cuts, given in Eq.~(\ref{eq:sel-cut}),  to calculate all the asymmetries for the polarizations and spin correlations discussed in Section~\ref{sec:spincorr-formlaism} for a set of anomalous couplings in order to obtain a semi-analytical expression for the observables. We use twenty six such benchmark couplings (five linear for each $\Delta C_1$, four linear for each $C_2$, four for $\Delta C_1^A$-$\Delta C_1^V$ cross terms, and four for $C_2^A$-$C_2^V$ cross terms) to obtain the semi-analytical expressions for the cross sections in four bins (Eq.~(\ref{eq:sigma-bin}) and all asymmetries. For the cross sections, the following expression is used to fit the data~\cite{Rahaman:2019lab}:
	\begin{equation}\label{eq:fit-cs}
		\sigma(\{C_i\})=\sigma_{\text{SM}} + \sum_i C_i \sigma_i + \sum_{i,j} C_iC_j\sigma_{ij}
	\end{equation}
	with 
	$ \{C_i\} =\left\{ \Delta C_1^A,\Delta C_1^V, C_2^A,C_2^V \right\}$,
	$\sigma_i$ as the linear/interference terms for couplings $C_i$, and $\sigma_{ij}$ as the quadratic or cross terms for couplings $C_i$ and $C_j$.
	The numerators of the asymmetries ($\Delta\sigma$) are fitted separately using the same form of the expression in Eq.~(\ref{eq:fit-cs}) and used in the asymmetries as
	\begin{equation}
		{\cal A}_j(\{C_i\}) = \dfrac{\Delta\sigma_{{\cal A}_j}(\{C_i\})}{\sigma(\{C_i\})}.
	\end{equation} 
	With the obtained semi-analytical expression for the observables in hand, we study the sensitivity of all observables to the anomalous couplings by varying one parameter at a time 
		in terms of $\chi^2$. 
	%for various combinations of polarizations and spin correlations for all four anomalous couplings. 
	The $\chi^2$ for a coupling $C_i$ is defined as follows,
	\begin{equation}\label{eq:def-chisq}
		\chi^2(C_i) =\sum_{n=1}^{N} \left|\dfrac{{\cal O}_n(C_i)-{\cal O}_n(C_i=0)}{\delta{\cal O}_n}\right|^2,
	\end{equation}
	where $N$ is the total number of observable ${\cal O}$, and  $\delta{\cal O}$ is the estimated error in ${\cal O}$. The error for the cross section ($\sigma$) and asymmetries (${\cal A}_i$) are
	\begin{align}
		\delta\sigma_i=\sqrt{\dfrac{\sigma_i}{{\cal L}} + (\epsilon_{\sigma_i}\sigma_i)^2}~~~~\text{and}~~~
		\delta {\cal A}_i=\sqrt{\dfrac{1-{\cal A}_i^2}{{\cal L}\times \sigma}+\epsilon_{\cal A}^2} ,
	\end{align}
	respectively with  ${\cal L}$ as the integrated luminosity;  $\epsilon_\sigma$ and $\epsilon_{\cal A}$ are the systematic uncertainty for the cross section and the asymmetries, respectively. 
We assume a flat systematic uncertainty of $10\%$ for the cross sections, i.e.,  $\epsilon_\sigma=0.1$~\cite{CMS:2019too} for all four bins in the four-lepton channel. For asymmetries, we use an absolute uncertainty of  $\epsilon_{\cal A}=0.01$~\cite{CMS:2018jcg} as a conservative choice.

	The sensitivity of polarization, spin correlation parameters, and their combinations are studied in terms of  $\chi^2$, and they are shown in Fig.~\ref{fig:bsm-asym}.
	The polarizations of top quarks ($Pol\l[t,\bar{t}\r]$) give higher $\chi^2$ value only in the negative side  for the couplings $\Delta C_1^V$, while for $\Delta C_1^A$, $Pol\l[t,\bar{t}\r]$ give higher  $\chi^2$ value only in the positive side, see Fig.~\ref{fig:bsm-asym} {\em top-row}. Polarization of $Z$ ($Pol[Z]$) on the other hand, give higher $\chi^2$ (comparable to $Pol\l[t,\bar{t}\r]$)   in the positive and negative side of $\Delta C_1^V$ and $\Delta C_1^A$, respectively. In case of $C_2^{A/V}$ ({\em bottom-panel} Fig.~\ref{fig:bsm-asym}), $Pol[Z]$ dictates the $\chi^2$ symmetrically compared to the $Pol\l[t,\bar{t}\r]$. These behavior are also seen in the angular distribution of polarization variables with benchmark anomalous couplings shown in Fig.~\ref{fig:dist-bsm-pol}. Besides this, the $\chi^2$ for the $\Delta C_1^A$ creates a dip in the positive side mainly due to $Pol[Z]$; this would result in a two patches interval in the limit. The spin correlation asymmetries, although not significant compared to polarization individually, they become significant enough when added together ($Cor[All]$), shown in {\em dotted/green} lines in all four couplings. The three body spin correlations of $\ttz$ ($Cor[\ttz]$) have a visible effect for the $\Delta C_1^A$ couplings. The spin correlation asymmetries show a smaller effect compared to the polarization asymmetries for an obvious reason related to the relation of the asymmetries to the polarization and spin correlation parameters. There are extra factors in the spin correlation asymmetries compared to the polarization asymmetries (see Ref.~\cite{Rahaman:2021fcz} for details), reducing the sensitivity to anomalous couplings compared to polarization asymmetries.

	We now compute the $\chi^2$ for the cross sections combining all four bins ($XSec$), $XSec$ combined with the polarizations and spin correlations successively and show them in Fig.~\ref{fig:bsm-cs-polcor} by varying one parameter at a time for the comparison. The sensitivity or the $\chi^2$ for the cross section in the binned case is better compared to the un-binned cross section, see appendix~\ref{app:bin-vs-unbin}. The dashed horizontal lines at $\chi^2=3.84$ indicate the $95\%$ C.L. bound on the couplings. The polarizations play a crucial role compared to the cross section in constraining the limits on the couplings, particularly for the couplings $\Delta C_1^{V/A}$. The polarizations  improve the limits on the couplings better on the positive side when added to the cross section for both $\Delta C_1^{V}$ and $\Delta C_1^{A}$. The cross sections  create two patches for $\Delta C_1^{A}$  within $95\%$ C.L. limit; The dip in the right patch reduces when polarizations and correlations are added successively, see {\em right-top} panel in Fig.~\ref{fig:bsm-cs-polcor}. For the couplings $C_2^{V/A}$, the cross sections dominate in constraining them; the polarizations and correlations improve the limits a little on top of the cross section. 
		These behavior are further illustrated in the $95\%$ C.L. contours ($\chi^2=5.991$~\cite{Cowan:2010js}) in $\Delta C_1^V$-$\Delta C_1^A$  and $C_2^V$-$C_2^A$ planes in Fig.~\ref{fig:bsm-cs-polcor}, {\em bottom-panel}. The cross section allows large values of simultaneous couplings for $\Delta C_1^V$ and $\Delta C_1^A$, e.g., $\Delta C_1^V=-1.2$ and $\Delta C_1^A=1.5$. This is due to a large cancellation of cross sections for $\Delta C_1^V$ and $\Delta C_1^A$. The polarizations reduce the allowed parameter space to two narrow regions  in the $\Delta C_1^A$ direction when added to the cross section, one of them includes the SM ($0,0$) point. The spin correlations further reduce the two regions of parameter space; the region not containing the SM point shrinks more, making less allowable region at  $95\%$ C.L. limit. In the $C_2^V$-$C_2^A$ plane, however, the regions are circular, and the polarizations and spin correlations have a comparatively smaller effect in shrinking the region compared to $\Delta C_1^V$-$\Delta C_1^A$ plane. We computed the one parameter $95\%$ C.L. limit on the couplings using cross sections, polarizations, and spin correlations for four sets of integrated luminosity,  such as  ${\cal L}=150$ fb$^{-1}$, $300$ fb$^{-1}$, $1000$ fb$^{-1}$, and $3000$ fb$^{-1}$,
		and listed them in Table~\ref{tab:limits-4l-oneparam}. We also obtained the limits on the operator's Wilson coefficients by changing the expression for the observable using relation given in Eq.~(\ref{eq:op-to-lag}) and listed them for the same set of luminosities in Table~\ref{tab:limits-4l-oneparam}. 
	%%%%%%%%%%%%%%%%%%%%%%%%%%%%%
	%%%%%%%%%%
	\begin{table}[h!]
		\caption{\label{tab:limits-4l-oneparam} One parameter $95\%$ C.L. limits on the  couplings in the $\ttz:4l$ channel at $\sqrt{s}=13$ TeV and ${\cal L}=150$ fb$^{-1}$, $300$ fb$^{-1}$, $1000$ fb$^{-1}$, and $3000$ fb$^{-1}$.}
		\renewcommand{\arraystretch}{1.8}
		\centering
		{\scriptsize  
			\begin{tabular}{|c|c|c|c|c|}\hline
				Coupling & $150$ fb$^{-1}$ & $300$ fb$^{-1}$ & $1000$ fb$^{-1}$& $3000$ fb$^{-1}$\\	\hline		
				$\Delta C_1^V   $& $\in[-0.787  , +1.01 ] $& $\in[-0.660        , +0.873        ] $& $\in[-0.501        , +0.681        ] $& $\in[-0.397  , +0.507  ]$ \\ \hline
				$\Delta C_1^A   $& \begin{tabular}{c}$\in[-0.220  , +0.400] $\\$\in[+0.732,+1.34 ] $\end{tabular}& \begin{tabular}{c}$\in[-0.171        , +0.248        ] $\\$\in[+0.887          ,+1.288]$\end{tabular}& \begin{tabular}{c}$\in[-0.116        , +0.145        ] $\\$\in[+1.00         ,+1.226]$\end{tabular}& \begin{tabular}{c}$\in[-0.0896 , +0.105  ]$\\ $\in[+1.077,  +1.170] $\end{tabular} \\ \hline
				
				$C_2^V          $& $\in[-0.101  , +0.101] $& $\in[-0.0860       , +0.086        ] $& $\in[-0.0664       , +0.0665       ] $& $\in[-0.0540 , +0.0539 ]$ \\ \hline
				$C_2^A          $& $\in[-0.1062 , +0.109] $& $\in[-0.0901       , +0.0926       ] $& $\in[-0.0688       , +0.0716       ] $& $\in[-0.055  , +0.0583 ]$ \\ \hline \hline
				$\frac{C_{\varphi t}  }{\Lambda^2}~ \left(\text{TeV}^{-2}\right) $& $\in[-10.831 , +2.553] $& $\in[-10.168       , +2.000        ] $& $\in[-8.503        , +-8.841       ] $& $\in[-1.356  , +1.063  ]$ \\ \hline
				$\frac{C_{\varphi Q}^-}{\Lambda^2}~ \left(\text{TeV}^{-2}\right) $& $\in[-3.401  , +5.137] $& $\in[-2.626        , +3.522        ] $& $\in[-1.793        , +2.145        ] $& $\in[-1.391  , +1.575  ]$ \\ \hline
				$\frac{C_{tZ}         }{\Lambda^2}~ \left(\text{TeV}^{-2}\right) $& $\in[-1.032  , +1.055] $& $\in[-0.875        , +0.899        ] $& $\in[-0.669        , +0.696        ] $& $\in[-0.537  , +0.567  ]$ \\ \hline
				$\frac{C_{tZ}^I       }{\Lambda^2}~ \left(\text{TeV}^{-2}\right) $& $\in[-0.981  , +0.984] $& $\in[-0.835        , +0.838        ] $& $\in[-0.645        , +0.646        ] $& $\in[-0.524  , +0.523  ]$ \\ \hline
				
			\end{tabular}
		}
	\end{table}

\begin{figure}[t!]
		\begin{center}
			\includegraphics[width=0.46\textwidth]{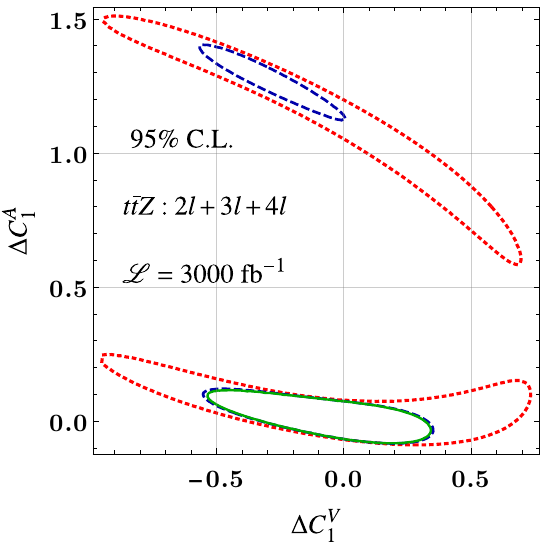}		
		\end{center}
		\caption{\label{fig:contr2l3l4l}  The two-parameter $95\%$ C.L. contours in $\Delta C_1^V$-$\Delta C_1^A$ 
				combining all the leptonic channel, i.e., $\ttz:2l+3l+4l$ channels  for $\sqrt{s}=13$ TeV and integrated luminosity ${\cal L}=3000$ fb$^{-1}$. The legends are the same as in Fig.~\ref{fig:bsm-cs-polcor}.   }
	\end{figure}

		For a complete analysis,  one could use all the leptonic channels of the $\ttz$ process providing better limits on the couplings.   All the polarizations and spin correlations can be obtained in different decay channels as follows:
	\begin{itemize}
		\item  $Z\to l^+l^-$,  $t/\bar{t}\to$ hadronic ($2l$) : Only the $Z$ polarizations are obtained in this case,
		\item $Z\to l^+l^-$,  $t\to$ leptonic, $\bar{t}$ hadronic ($3l$): Top quark polarizations are obtained along with the $t$-$Z$ spin correlations by reconstructing the missing neutrino~\cite{Rahaman:2019lab}. The $\bar{t}$ polarizations and $\tb$-$Z$ spin correlations are obtained by reversing the top and anti-top decay,
		\item Fully leptonic ($4l$): In this case, $t\bar{t}$ and $\ttz$ spin correlations are obtained.
	\end{itemize} 
	The hadronic $Z$-decays are avoided to reconstruct the $\ttz$ topology better. We estimated the cross section and the detection efficiency for the $2l$ and $3l$ topology in the SM and adjusted the statistics for all the polarizations and spin correlations in accordance with the above categorization.  
	The cross sections in $2l$ and $3l$ channels increase by a factor of $\sim9.9$ and $\sim 2.6$ from the $4l$ channel, respectively.
	We then estimate the one parameter projected limits at $95\%$ C.L. on the couplings  by combining the three channels, i.e., in the $2l+3l+4l$ channel for the set of luminosity of ${\cal L}=150$ fb$^{-1}$, $300$ fb$^{-1}$, $1000$ fb$^{-1}$, and $3000$ fb$^{-1}$ and listed them in Table~\ref{tab:limits-4l3l2l-single}. The limits in $2l+3l+4l$ channels are tighter roughly by a factor of $2$ compared to the $4l$ channel. We use systematic uncertainties of $0.13$ and $0.16$ for the cross sections in the $3l$ and $2l$ channels, respectively, and  the same uncertainty of $0.01$ for the asymmetries as used for the $4l$ channel. 
	%The limits on the operator, in this case, are now comparable to the existing limits obtained at CMS~\cite{CMS:2019too,CMS:2022hjj}. 
	We also redraw the $95\%$ C.L. contour for the $2l+3l+4l$ channel in $\Delta C_1^V$-$\Delta C_1^A$ plane at ${\cal L}=3000$ fb$^{-1}$ to see the improvement, see Fig.~\ref{fig:contr2l3l4l}. The parameter space is tighter compared to only the $4l$ channel, as depicted in Fig.~\ref{fig:bsm-cs-polcor}. Unlike the $4l$ channel, the cross sections divide the parameter space into two regions. The region not containing the SM point is smaller than the point containing the SM point when polarizations are added to the cross sections. The region not containing the SM point further disappears when spin correlations come into play, allowing the parameter space only in the neighborhood of the SM point.
	%%%%%%%%%%%%
	\begin{table}[h!]
		\caption{\label{tab:limits-4l3l2l-single}  One parameter $95\%$ C.L. limits on the  couplings projected in the $\ttz:4l+3l+2l$ channel at $\sqrt{s}=13$ TeV and ${\cal L}=150$ fb$^{-1}$, $300$ fb$^{-1}$, $1000$ fb$^{-1}$, and $3000$ fb$^{-1}$.}
		\renewcommand{\arraystretch}{1.8}
		\centering
		{\scriptsize  
			\begin{tabular}{|c|c|c|c|c|}\hline
				Coupling & $150$ fb$^{-1}$ & $300$ fb$^{-1}$ & $1000$ fb$^{-1}$& $3000$ fb$^{-1}$\\	\hline		
				$\Delta C_1^V   $& $\in[-0.469  , +0.643        ] $& $\in[-0.414        , +0.548        ] $& $\in[-0.335        , +0.382        ] $& $\in[-0.267        , +0.263        ]$ \\ \hline    
				$\Delta C_1^A   $&\begin{tabular}{c} $\in[-0.0988 , +0.119        ] $\\$\in[+1.015,+      1.22] $\end{tabular}&\begin{tabular}{c} $\in[-0.0833       , +0.097        ] $\\$\in[+1.043,+   1.202] $\end{tabular}&\begin{tabular}{c} $\in[-0.0648       , +0.0723       ] $\\$\in[+1.09,+    1.168] $\end{tabular}&$\in[-0.054        , +0.0587       ]$ \\ \hline
				$C_2^V          $& $\in[-0.0605 , +0.0634       ] $& $\in[-0.053        , +0.0561       ] $& $\in[-0.0427       , +0.0462       ] $& $\in[-0.0356       , +0.0394       ]$ \\ \hline                        $C_2^A          $& $\in[-0.0589 , +0.0588       ] $& $\in[-0.052        , +0.0518       ] $& $\in[-0.0425       , +0.0421       ] $& $\in[-0.036        , +0.0355       ]$ \\ \hline\hline
				$\frac{C_{\varphi t}  }{\Lambda^2}~ \left(\text{TeV}^{-2}\right) $& $\in[-1.606  , +1.182        ] $& $\in[-1.269        , +0.999        ] $& $\in[-0.913        , +0.777        ] $& $\in[-0.719        , +0.642        ]$ \\ \hline    
				$\frac{C_{\varphi Q}^-}{\Lambda^2}~ \left(\text{TeV}^{-2}\right) $& $\in[-1.525  , +1.778        ] $& $\in[-1.289        , +1.459        ] $& $\in[-1.01         , +1.103        ] $& $\in[-0.849        , +0.906        ]$ \\ \hline    
				$\frac{C_{tZ}         }{\Lambda^2}~ \left(\text{TeV}^{-2}\right) $& $\in[-0.588  , +0.617        ] $& $\in[-0.515        , +0.545        ] $& $\in[-0.415        , +0.449        ] $& $\in[-0.346        , +0.383        ]$ \\ \hline
				$\frac{C_{tZ}^I       }{\Lambda^2}~ \left(\text{TeV}^{-2}\right) $& $\in[-0.572  , +0.572        ] $& $\in[-0.505        , +0.504        ] $& $\in[-0.413        , +0.410        ] $& $\in[-0.35         , +0.345        ]$ \\ \hline			
			\end{tabular}
		}
	\end{table}
	%%%%%%%%%%%%%%%

	%%%%%%%%%%%%%%%%%%%%
	\begin{figure}[t!]
		\begin{center}
			\includegraphics[width=0.47\textwidth]{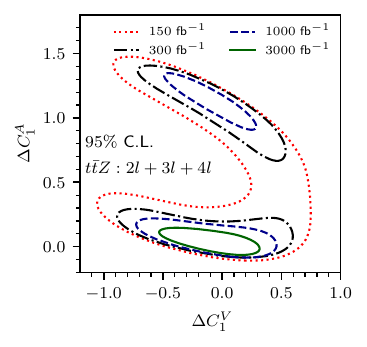}
			\includegraphics[width=0.5\textwidth]{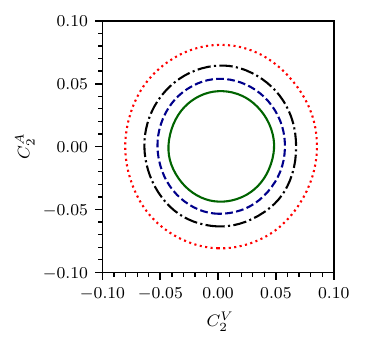}
			\includegraphics[width=0.49\textwidth]{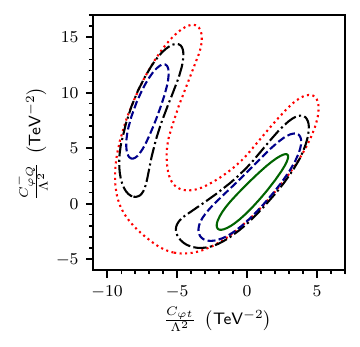}
			\includegraphics[width=0.49\textwidth]{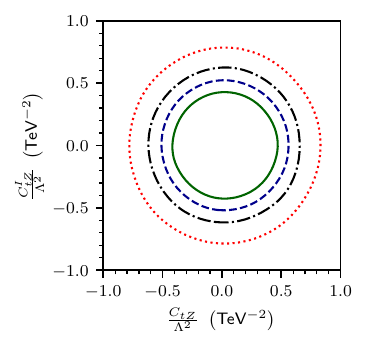}
		\end{center}
		\caption{\label{fig:Lag-comp-Lum} The $95\%$ C.L. BCI contours from MCMC are shown
				for the couplings of the effective Lagrangian in Eq.~(\ref{eq:BSM-lag}) (in {\em top-panel}) and their translated contours to the couplings of the effective operators in Eq.~(\ref{eq:BSM-Op}) (in {\em bottom-panel}) in $\ttz:2l+3l+4l$ channels for  various luminosities at $\sqrt{s}=13$ TeV using  the cross section and all asymmetries. }
	\end{figure}
	\begin{figure}[t!]
		\begin{center}
			\includegraphics[width=1\textwidth]{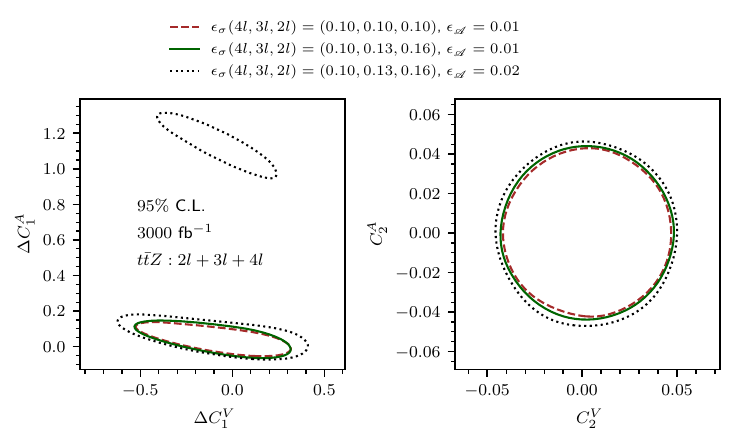}
		\end{center}
		\caption{\label{fig:Lag-comp-eps} The $95\%$ C.L. BCI contours from MCMC in $\Delta C_1^V$-$\Delta C_1^A$ ({\em left-panel}) and $C_2^V$-$C_2^A$ ({\em right-panel}) planes are shown in $\ttz:2l+3l+4l$ channels
				for the couplings of the effective Lagrangian in Eq.~(\ref{eq:BSM-lag})  for  various combination of systematic uncertainties with ${\cal L}=3000$ fb$^{-1}$ at $\sqrt{s}=13$ TeV using  the cross section and all asymmetries.   } 
	\end{figure}
	%%%%%%%%%%%%%%%%%%%%%%%%%%%%%
	%%%%%%%%%%
	
	\begin{table}[h!]
		\caption{\label{tab:limits-2l3l4l-mcmc} Simultaneous BCI limits at  $95\%$  C.L.  on the  couplings are listed  projected in the $\ttz:2l+3l+4l$ channel at $\sqrt{s}=13$ TeV and ${\cal L}=150$ fb$^{-1}$, $300$ fb$^{-1}$, $1000$ fb$^{-1}$, and $3000$ fb$^{-1}$.}
		\renewcommand{\arraystretch}{1.8}
		\centering
		{\footnotesize  
			\begin{tabular}{|c|c|c|c|c|}\hline
				Coupling & $150$ fb$^{-1}$ & $300$ fb$^{-1}$ & $1000$ fb$^{-1}$& $3000$ fb$^{-1}$\\	\hline		
				$ \Delta C_1^V   $& $\in[-0.814 , + 0.642 ] $& $\in[-0.716 , + 0.526 ] $& $\in[-0.543 , + 0.331 ] $& $\in[-0.444 , +0.244]$ \\ \hline
				$ \Delta C_1^A   $& $\in[-0.074 , + 1.402 ] $& $\in[-0.071 , + 1.379 ] $& $\in[-0.052 , + 1.323 ] $& $\in[-0.042 , +0.129]$ \\ \hline
				$ C_2^V          $& $\in[-0.067 , + 0.072 ] $& $\in[-0.055 , + 0.061 ] $& $\in[-0.042 , + 0.048 ] $& $\in[-0.035 , +0.041]$ \\ \hline
				$ C_2^A          $& $\in[-0.068 , + 0.068 ] $& $\in[-0.058 , + 0.057 ] $& $\in[-0.044 , + 0.044 ] $& $\in[-0.036 , +0.037]$ \\ \hline\hline
				$ \frac{C_{\varphi t} }{\Lambda^2}~ \left(\text{TeV}^{-2}\right) $& $\in[-8.739 , + 3.642 ] $& $\in[-8.758 , + 3.597 ] $& $\in[-8.791 , + 3.137 ] $& $\in[-1.729 , +2.410]$ \\ \hline
				$ \frac{C_{\varphi Q}^-}{\Lambda^2}~ \left(\text{TeV}^{-2}\right)$& $\in[-3.464 , +13.145 ] $& $\in[-3.188 , +12.104 ] $& $\in[-2.425 , +11.620 ] $& $\in[-1.825 , +3.672]$ \\ \hline
				$ \frac{C_{tZ} }{\Lambda^2}~ \left(\text{TeV}^{-2}\right)        $& $\in[-0.654 , + 0.700 ] $& $\in[-0.539 , + 0.592 ] $& $\in[-0.412 , + 0.469 ] $& $\in[-0.343 , +0.394]$ \\ \hline
				$ \frac{C_{tZ}^I }{\Lambda^2}~ \left(\text{TeV}^{-2}\right)      $& $\in[-0.661 , + 0.660 ] $& $\in[-0.563 , + 0.556 ] $& $\in[-0.425 , + 0.431 ] $& $\in[-0.354 , +0.357]$ \\ \hline
			\end{tabular}
		}
	\end{table}
		\paragraph{Simultaneous limits : }
		We have estimated the limits above by varying one parameter at a time, fixing others to their SM values. More realistic bounds on the couplings can be obtained by simultaneously varying all the parameters followed by marginalizing them. Here,  we extract simultaneous limits on the couplings using the Markov-Chain--Monte-Carlo (MCMC) method and {\tt GetDist}~\cite{Lewis:2019xzd} package for marginalization. The simultaneous $95\%$ C.L. limits in $4l+3l+2l$ channel are shown in Table~\ref{tab:limits-2l3l4l-mcmc} for the same set of luminosity ${\cal L}=150$ fb$^{-1}$, $300$ fb$^{-1}$, $1000$ fb$^{-1}$, and $3000$ fb$^{-1}$. The simultaneous limits are diluted by roughly a factor of $2$ for almost all the couplings compared to the one parameter limits (Table~\ref{tab:limits-4l3l2l-single}). The corresponding marginalized  Bayesian Credible Interval (BCI) at $95\%$ C.L. are shown for the benchmark luminosities in Fig.~\ref{fig:Lag-comp-Lum} in the  $\Delta C_1^V$-$\Delta C_1^A$  and $C_2^V$-$C_2^A$ planes. As the luminosity increases to $300$ fb$^{-1}$ from $150$ fb$^{-1}$, the region of parameter space get divided along $\Delta C_1^A$ direction in $\Delta C_1^V$-$\Delta C_1^A$ plane. The two region shrinks at $1000$ fb$^{-1}$ further, and finally, we only have one region centering SM point at  $3000$ fb$^{-1}$. This behavior is translated to the $C_{\varphi t}$-$C_{\varphi Q}^-$ plane. The contours in the $C_2^V$-$C_2^A$ plane ($C_{tZ}$-$C_{tZ}^I$ plane) shrink as the luminosity increases remaining circular with the SM point at the center.

		\paragraph{Effect of systematic uncertainties : }
		The simultaneous limits change due to changes in systematic uncertainties. Increasing the systematic uncertainties loosen the limits on the couplings. A comparison of the limits is shown in $95\%$ C.L. BCI contours in Fig.~\ref{fig:Lag-comp-eps}  for different combinations of systematic uncertainties for cross sections and asymmetries at $3000$ fb$^{-1}$ luminosity. The contours ({\em dashed}/red lines) get tightened up by little when systematic uncertainties for cross sections are considered to be $0.1$ for all $4l$, $3l$ and $2l$ channel keeping $\epsilon_{\cal A}=0.01$  as compared to the benchmark systematic uncertainties of $\epsilon_\sigma(4l,3l,2l)=(0.10,0.13,0.16),~\epsilon_{\cal A}=0.01$ ({\em solid}/green lines) in both  $\Delta C_1^V$-$\Delta C_1^A$  and $C_2^V$-$C_2^A$ planes. The $\epsilon_{\cal A}$ has comparatively large effects on the limits; The contours enlarge for $\epsilon_{\cal A}=0.02$ keeping the $\epsilon_\sigma$s to their benchmark values. A second region way from the SM point appears along the $\Delta C_1^A$ direction in the $\Delta C_1^V$-$\Delta C_1^A$ plane, drastically changing the limits on $\Delta C_1^A$.

		The final  one parameter  limits that we estimated in $4l+3l+2l$ channels on the operators' couplings for ${\cal L}=150$ fb$^{-1}$ (second-column, Table~\ref{tab:limits-4l3l2l-single})  are better compared to observed one parameter limits  by CMS  with ${\cal L}=77.5$ fb$^{-1}$~\cite{CMS:2019too}  and ${\cal L}=138$ fb$^{-1}$~\cite{CMS:2022hjj}. Our  simultaneous limits
		for ${\cal L}=150$ fb$^{-1}$ (second-column, Table~\ref{tab:limits-2l3l4l-mcmc})  are also better compared to the simultaneous limits observed by CMS~\cite{CMS:2022hjj}. 
		Our one parameter limits on the operator for ${\cal L}=150$ fb$^{-1}$  are better in contrast to the one parameter limits obtained by SMFiT~\cite{Hartland:2019bjb} collaboration in a global analysis of the top quark sector. Our simultaneous limits are also better, except $C_{\varphi Q}^-$, compared to the marginalized limits obtained by SMFiT~\cite{Hartland:2019bjb}.

	We did not account for backgrounds other than the SM $\ttz$ process but used a large systematic uncertainty ($10\%$, $13\%$ and $16\%$ for $4l$, $3l$ and $2l$ channel) in our analysis. Further, we use the default isolation criteria in the {\tt Delphes}, which are relatively strict compared to what is used in CMS analyses giving us fairly conservative results. 
		We note that the authors in Ref.~\cite{Ravina:2021kpr} also study anomalous couplings in $t\bar{t}Z$ production process, where  the polarization of top quarks and spin correlations of top and anti-top pair are estimated in a truth level simulation. In contrast to Ref.~\cite{Ravina:2021kpr}, we estimate polarization of $Z$ boson including the top quarks, spin correlation of top and anti-top system, top and $Z$ system as well as the top, anti-top, and $Z$ system, i.e., all possible two-body and three-body spin correlations, by reconstructing the top quarks in a detailed detector level simulation. The marginalized constraints on the operator, including a $CP$-odd operator,  in our study are  tighter as compared to what was obtained in  Ref.~\cite{Ravina:2021kpr}, where {\bf Run 2} and  {\bf Run 2}+{\bf Run 3} limits can be compared with our limits for ${\cal L}=150$ fb$^{-1}$ and $300$ fb$^{-1}$, respectively.
	
	%%%%%%%%%%%%%%%%%%%%%%%%%%%%%%
	\section{Summary}\label{sec:conclusion}
	In summary, we studied the anomalous $\ttz$ interaction in leptonic channel of the $\ttz$ production process at the $13$ TeV LHC with the help of polarizations, two-body and three-body spin correlations of $t$, $\tb$ and $Z$ on top of the  cross sections binned in $p_T(Z)$.
	We showed how the reconstruction of two neutrinos at the detector level affects the angular distributions corresponding to polarizations and spin correlations compared to the parton level distributions. We identified a few polarization and spin correlation parameters sensitive to only one kind of anomalous couplings, helping us disentangle the effect of the four anomalous couplings that we have considered. The sensitivity of the couplings to polarizations and spin correlations are studied in the form of $\chi^2$ for a luminosity of ${\cal L}=3000$ fb$^{-1}$. The improvements of limits on the couplings are studied over the binned cross sections by successively including the polarization and spin correlation asymmetries. We estimated the one parameter and simultaneous limits at $95\%$ C.L.  on the anomalous couplings as well as the effective operators for a set of luminosities of $150$ fb$^{-1}$, $300$ fb$^{-1}$, $1000$ fb$^{-1}$, and $3000$ fb$^{-1}$.
	Our limits on the couplings (except $\Delta C_1^A$/ $C_{\varphi t}$) are  better compared to the existing limits obtained by CSM~\cite{CMS:2019too,CMS:2022hjj}.
	The polarizations and spin correlations  help in tightening the region of parameters space, especially for the vector and axial-vector couplings ($\Delta C_1^V$-$\Delta C_1^A$ ) in comparison to the cross section by a considerable amount. The parameter space becomes even tinier in the $2l+3l+4l$ channel neighboring only the SM point. Our strategy in this analysis can serve as an extra handle in interpreting anomalous interactions on the data at the high energy and high luminosity LHC.
	
	%%%%%%%%%%%%%%%%%
	\paragraph*{ACKNOWLEDGEMENT}
	The author would like to acknowledge support from the Department of Atomic Energy, Government of India, for the Regional Centre for Accelerator-based Particle Physics (RECAPP), Harish Chandra Research Institute.
	
	\newpage
	\appendix
	\section{Standard model values of polarizations, spin correlations, and their asymmetries}\label{app:SM-pol-asym}
	\begin{table}[h!]
		\caption{\label{tab:sm-pol-cor} Standard Model values of various polarizations and spin correlations along with their asymmetries above in parton level as well as in detector level in the $\ttz$ production in the $4l$ channel at the $\sqrt{s}=13$ TeV LHC. The values are listed with asymmetries more than  $5\sigma$ MC error of $10$ million events ($\delta {\cal A} =0.001$) in {\tt Delphes} level. 
			PT, PR and DR stands for  {\tt Parton-Truth}, {\tt Parton-Reco}, and {\tt Delphes-Reco}, respectively.	 
		} 	
		\renewcommand{\arraystretch}{1.49}
		\begin{tabular}{|c|ccc|cccc|}\hline
			&\multicolumn{3}{|c|}{Asymmetry}  &\multicolumn{4}{c|}{ Values for Pol. and Corr.}  \\ \hline
			\text{Param} &  \text{PT} & \text{PR} & \text{DR}& \text{PT} & \text{PR} & \text{DR}& \text{DR-Error}  \\\hline 
			$p_x^t$ & 0.0327 & 0.0572 & 0.0575 & -0.165 & -0.289 & -0.291  & 0.00518 \\
			$p_z^t$ & -0.0992 & -0.0525 & -0.0737 & 0.501 & 0.265 & 0.372  & 0.00518 \\
			$p_x^{\bar{t}}$ & -0.00155 & 0.028 & 0.0332 & 0.00782 & -0.142 & -0.168  & 0.00519 \\
			$p_z^{\bar{t}}$ & -0.0705 & -0.0274 & -0.0584 & 0.356 & 0.138 & 0.295  & 0.00518 \\
			$T_{xz}^Z$ & 0.00606 & 0.00619 & -0.0191 & 0.0117 & 0.0119 & -0.0368  & 0.00198 \\
			$T_{x^2-y^2}^Z$ & -0.106 & -0.106 & -0.144 & -0.407 & -0.407 & -0.556  & 0.00391 \\
			$T_{zz}^Z$ & -0.079 & -0.079 & -0.131 & -0.172 & -0.172 & -0.285  & 0.00222 \\
			$pp_{xx}^{t\bar{t}}$ & -0.0209 & -0.0179 & -0.0202 & -0.532 & -0.458 & -0.515  & 0.0262 \\
			$pp_{xz}^{t\bar{t}}$ & 0.0134 & 0.00557 & -0.0188 & 0.342 & 0.142 & -0.479  & 0.0262 \\
			$pp_{yy}^{t\bar{t}}$ & 0.0116 & 0.0124 & -0.011 & 0.295 & 0.317 & -0.281  & 0.0262 \\
			$pp_{zx}^{t\bar{t}}$ & 0.0144 & 0.00229 & -0.0229 & 0.367 & 0.0585 & -0.584  & 0.0262 \\
			$pp_{zz}^{t\bar{t}}$ & 0.00679 & -0.00488 & -0.0279 & 0.173 & -0.125 & -0.713  & 0.0262 \\
			$pT_{x(x^2-y^2)}^{tZ}$ & -0.00688 & -0.00915 & -0.0118 & 0.164 & 0.218 & 0.28  & 0.0245 \\
			$pT_{x(zz)}^{tZ}$ & -0.00321 & -0.00528 & -0.00786 & 0.0432 & 0.0711 & 0.106  & 0.0138 \\
			$pT_{z(x^2-y^2)}^{tZ}$ & 0.0141 & 0.00901 & 0.0143 & -0.337 & -0.214 & -0.339  & 0.0245 \\
			$pT_{x(zz)}^{\bar{t}Z}$ & -0.00413 & -0.00592 & -0.00683 & 0.0556 & 0.0798 & 0.0919  & 0.0138 \\
			$pT_{z(x^2-y^2)}^{\bar{t}Z}$ & 0.00696 & 0.00263 & 0.00867 & -0.166 & -0.0626 & -0.206  & 0.0245 \\
			$ppT_{xz(zz)}^{t\bar{t}Z}$ & 0.00207 & 0.00221 & 0.00521 & 0.141 & 0.15 & 0.354  & 0.07 \\
			$ppT_{yy(x^2-y^2)}^{t\bar{t}Z}$ & 0.00213 & 0.00159 & 0.00617 & 0.256 & 0.192 & 0.742  & 0.124 \\
			$ppT_{zz(zz)}^{t\bar{t}Z}$ & 0.00731 & 0.00688 & 0.00961 & 0.498 & 0.468 & 0.654  & 0.0699 \\ \hline
		\end{tabular}
	\end{table}

		\section{Binned cross sections versus the un-binned cross section}\label{app:bin-vs-unbin}
\begin{figure}[ht!]
		\begin{center}
			\includegraphics[width=0.8\textwidth]{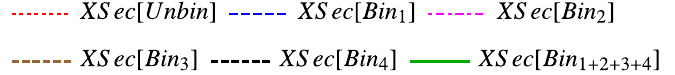}
			\includegraphics[width=0.49\textwidth]{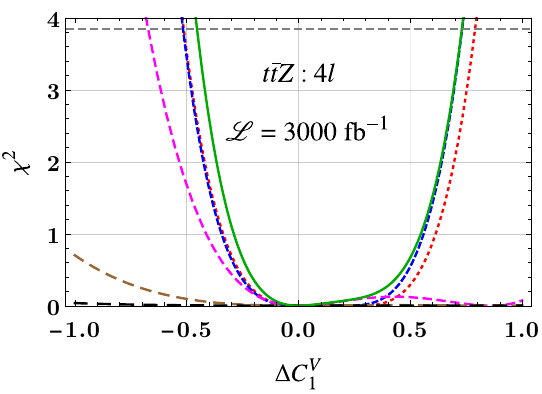}
			\includegraphics[width=0.49\textwidth]{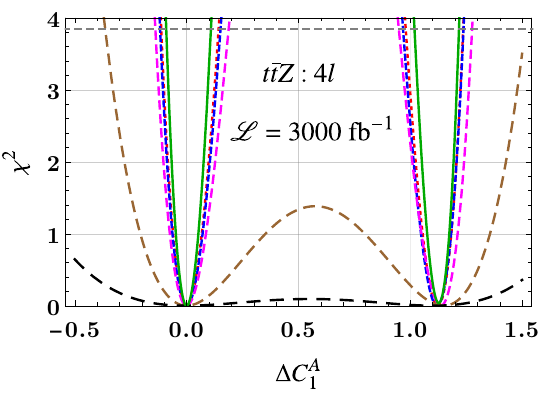}
			\includegraphics[width=0.49\textwidth]{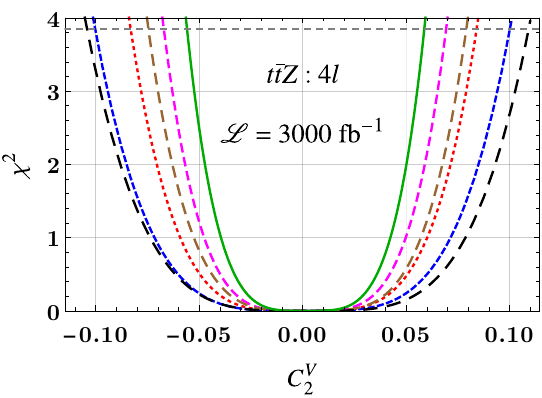}
			\includegraphics[width=0.49\textwidth]{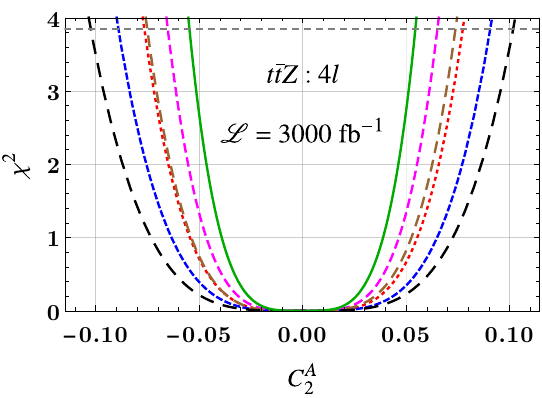}			
		\end{center}
		\caption{\label{fig:bsm-cs-bin}  Comparison of the total un-binned cross section with the binned cross sections (binned in $p_T(Z)$, see Eq.~(\ref{eq:sigma-bin})) in terms of $\chi^2$ as a function of $\ttz$ anomalous couplings at $\sqrt{s}=13$ TeV and integrated luminosity of  ${\cal L}=3000$ fb$^{-1}$.  } 
	\end{figure}

		Here, we compare the cross sections binned  over reconstructed $p_T(Z)$ with the total un-binned cross sections in terms of $\chi^2$ as a function of anomalous couplings, shown in Fig.~\ref{fig:bsm-cs-bin}.  
		The  combined binned cross section ({\em solid}/green lines)  performs better compared to the total un-binned cross section ({\em dotted}/red lines) in constraining the couplings. 
		Some individual bins even perform better compared to the total un-binned cross sections. For example, $Bin_1$ (blue lines) for $\Delta C_1^{V/A}$; and $Bin_2$ (magenta lines ) together with $Bin_3$ (brown lines) for $C_2^{V/A}$ provide tighter limits as compared to the total un-binned cross sections.
	
	%%%%%%%%%%%%%%%%%%%%%%%%%%%%%%%%%%%%%
	
	\newpage
	\bibliography{References}
	\bibliographystyle{utphysM}
	%%%%%%%%%%%%%%%%%%%%%%%%%%%%%%%%%%%%%%
\end{document}